\documentclass[twocolumn,superscriptaddress,reprint,preprintnumbers,amsmath,amssymb,aps,prb,floatfix,longbibliography,]{revtex4-2}

\usepackage{graphicx}% Include figure files
\usepackage[export]{adjustbox}
\usepackage{dcolumn}% Align table columns on decimal point
\usepackage{bm}% bold math
\usepackage{hyperref}
\usepackage{amsmath,amssymb}
\hypersetup{colorlinks=true, linkcolor=blue, citecolor=blue, filecolor=blue, urlcolor=blue}
\usepackage[normalem]{ulem}

\begin{document}
	
	\title{Manifestation of incoherent-coherent crossover and non-Stoner magnetism in the electronic structure of Fe$_3$GeTe$_2$}
	
	\author{Deepali Sharma}
	%\email{deepali19@iiserb.ac.in}
	\affiliation{Department of Physics, Indian Institute of Science Education and Research Bhopal, Bhopal Bypass Road, Bhauri, Bhopal 462066, India}%
	
	\author{Asif Ali}
	\affiliation{Department of Physics, Indian Institute of Science Education and Research Bhopal, Bhopal Bypass Road, Bhauri, Bhopal 462066, India}%
	
	\author{Neeraj Bhatt}
	\affiliation{Department of Physics, Indian Institute of Science Education and Research Bhopal, Bhopal Bypass Road, Bhauri, Bhopal 462066, India}%
	
	\author{Rajeswari Roy Chowdhury}
	\affiliation{Department of Physics, Indian Institute of Science Education and Research Bhopal, Bhopal Bypass Road, Bhauri, Bhopal 462066, India}%
	
	\author{Chandan Patra}
	\affiliation{Department of Physics, Indian Institute of Science Education and Research Bhopal, Bhopal Bypass Road, Bhauri, Bhopal 462066, India}%
	
	\author{Ravi Prakash Singh}
	\affiliation{Department of Physics, Indian Institute of Science Education and Research Bhopal, Bhopal Bypass Road, Bhauri, Bhopal 462066, India}%
	
	\author{Ravi Shankar Singh}
	\email{rssingh@iiserb.ac.in}
	\affiliation{Department of Physics, Indian Institute of Science Education and Research Bhopal, Bhopal Bypass Road, Bhauri, Bhopal 462066, India}%
	
	\date{\today}% It is always \today, today,
	
	\begin{abstract}
		
		Two-dimensional (2D) van der Waals ferromagnets have
		potential applications as next-generation spintronic devices and provide a platform to explore the fundamental physics behind 2D magnetism. The dual nature (localized and itinerant) of electrons adds further complexity to the understanding of correlated magnetic materials. Here, we present the temperature evolution of electronic structure in 2D van der Waals ferromagnet, Fe$_{3}$GeTe$_{2}$, using photoemission spectroscopy in conjunction with density functional theory (DFT) plus dynamical mean field theory (DMFT). With the appearance of quasiparticle peak and its evolution in the vicinity of Fermi energy, we unveil empirical evidences of incoherent-coherent crossover at around 125 K. DFT+DMFT results show that the quasiparticle lifetime surpasses thermal energy for temperature below 150 K, confirming incoherent-coherent crossover in the system. No appreciable change in the Fe 2\textit{p} core level, overall valence band spectra across the magnetic transition, and temperature dependent ferromagnetic DFT+DMFT results, provide substantial evidence for non-stoner magnetism in Fe$_{3}$GeTe$_{2}$. We elucidate the temperature dependent intimate relation between magnetism and electronic structure in Fe$_{3}$GeTe$_{2}$. Sommerfeld coefficient of $\sim$ 104 mJ mol$^{-1}$ K$^{-2}$ obtained in the low temperature limit from DFT+DMFT calculations resolve the long standing issue of large Sommerfeld coefficient ($\sim$ 110 mJ mol$^{-1}$ K$^{-2}$) obtained from specific heat measurements.
		
	\end{abstract}
	
	\maketitle
	
	\section{introduction}
	Magnetism in two-dimensional (2D) van der Waals (vdW) materials have been the subject of great interest for their ordered magnetic phases down to the monolayer limit  \cite{mak_probing_2019,gibertini_magnetic_2019,burch_magnetism_2018}. The magnetic and electronic ground state of these 2D materials can be manipulated by external stimuli such as strain, gating, proximity effects, $etc.$, and the easy exfoliation allows the fabrication of novel devices down to the 2D limit \cite{gibertini_magnetic_2019,burch_magnetism_2018}. In this class of ferromagnets, Cr$_{2}$Ge$_{2}$Te$_{6}$, CrSiTe$_{3}$, CrI$_{3}$, $etc.$ have been widely studied due to the rich interplay between long-range magnetic ordering, inter-site exchange (\textit{J}), and intra-site coulomb (\textit{U}) interactions in deciding the electronic structure \cite{gong_discovery_2017,williams_magnetic_2015,huang_layer-dependent_2017}.
	
	Fe$_{3}$GeTe$_{2}$, a 2D vdW ferromagnet, has gained enormous attention due to the remarkably high Curie temperature ($T_{C}$ $\sim$ 220 K), large uniaxial magneto-crystalline anisotropy persisting down to the monolayer limit, magnetic skyrmions, anomalous Hall effect, $etc.$ \cite{deng_gate-tunable_2018,birch_history-dependent_2022-2,chowdhury_unconventional_2021,roy_chowdhury_modification_2022}. Density functional theory (DFT) reveals the itinerant ferromagnetism fulfilling the Stoner criteria \cite{zhuang_strong_2016,yuan_tuning_2017}, which is supported by photoemission spectroscopy (PES) exhibiting continuous spectral weight transfer within Fe 3\textit{d} states (signature of exchange splitting) in the ferromagnetic phase \cite{zhang_emergence_2018}. However, in complete contrast, another report using angle-resolved PES (ARPES) reveals insignificant change in the band dispersion with increasing temperature up to $T_{C}$ \cite{xu_signature_2020}.
	
	In general, the nature of ferromagnetism is understood from [i] Stoner model in the case of itinerant bands where the temperature dependent exchange splitting of dispersive spin bands drive the long-range magnetic ordering, which eventually vanishes upon reaching $T_{C}$ or [ii] spin mixing model in the case of localized bands where the exchange splitting exists even above $T_{C}$ and the thermal fluctuation of the local moment reduces the magnetization \cite{santiago_itinerant_2017, korenman_local-band_1977, maiti_finite_2002}. However, many magnetic materials fall under the intermediate regime of itinerant-local moments, such as cuprates and iron-pnictides, due to a rich interplay of electronic states and magnetic correlations \cite{santiago_itinerant_2017,korenman_local-band_1977, maiti_finite_2002,johnston_puzzle_2010,Hansmann_LaFeAsODMFT_PRL2010,dai_magnetism_2012,haule_coherenceincoherence_2009}. Despite being a $d$-electron system, coexisting localized as well as itinerant electrons drive Fe$_{3}$GeTe$_{2}$ to a heavy-fermionic state at low temperature \cite{AFM,zhang_emergence_2018,zhao_kondo_2021}. The electronic transport and magnetic measurements indicate signature of incoherent-coherent crossover much below $T_{C}$, along with Fano-resonance feature in the scanning tunneling spectra concluding the Kondo scenario \cite{zhang_emergence_2018,rana_spin-polarized_2022,zhao_kondo_2021}. The large effective mass ($\sim$ 13.3 $m_{\text{DFT}}$) from Sommerfeld coefficient ($\gamma$ = 110 mJ mol$^{-1}$ K$^{-2}$) obtained from specific heat measurements is not adequately reproduced from ARPES and dynamical mean field theory (DMFT) calculations across the literature \cite{zhu_electronic_2016,zhang_emergence_2018,xu_signature_2020,kim_large_2018}. Understanding the nature of ferromagnetism (Stoner versus non-Stoner), incoherent-coherent crossover, and large effective mass leading to heavy-fermionic behavior warrants a comprehensive study of the electronic structure and its relation with magnetism in both high and low temperature limits.
	
	\begin{figure*}[t]
		\centerline{\includegraphics[width=0.85\textwidth]{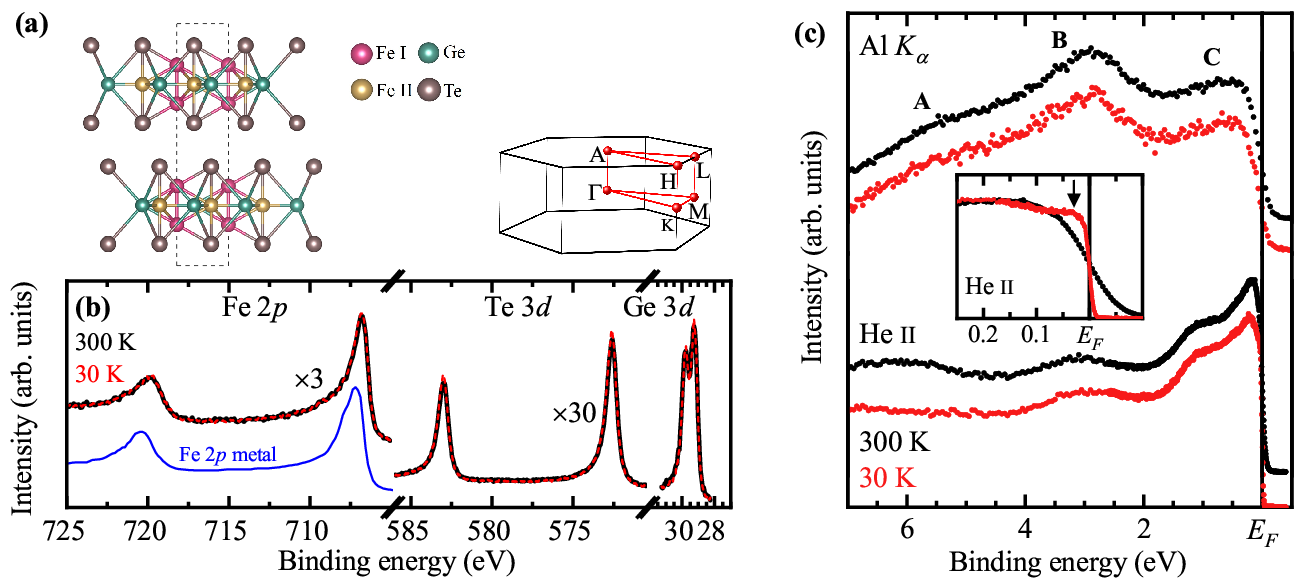}}
		\caption{(a) Crystal structure and Brillouin zone of Fe$_{3}$GeTe$_{2}$. The dashed line in the structure shows the unit cell comprising two formula units. (b) Core level photoemission spectra of Fe 2$p$, Te 3$d$ and Ge 3$d$ collected at at 300 K (black) and 30 K (red) using Al $K_{\alpha}$ (intensities are normalized to that of the survey scan, Fig. S1 of SM \cite{SM}). The spectra in blue line represent 2\textit{p} core level of metal Fe in ferromagnetic regime, reproduced from \cite{PhysRevB.55.11488}. (c) Valence band photoemission spectra collected using Al $K_{\alpha}$ and He {\scriptsize II} radiations at 300 K (black) and 30 K (red). Inset shows He {\scriptsize II} spectra in the vicinity of $E_{F}$.}\label{fig:fig1}
	\end{figure*} 
	
	Here, we investigate the electronic structure of Fe$_{3}$GeTe$_{2}$ using photoemission spectroscopy and theoretical calculations within DFT+DMFT framework. The Fe 2\textit{p} core level spectra as well as valence band spectra remain very similar across $T_C$, except for the states close to the Fermi level ($E_{F}$). High-resolution spectra unveils the emergence of a quasiparticle peak in the close vicinity of $E_{F}$ in the magnetically  ordered  phase. The overall valence band spectra and evolution of quasiparticle peak, and manifestation of incoherent-coherent crossover ($\sim$ 125 K) in the experimental spectra, are very well captured within temperature dependent ferromagnetic DFT+DMFT calculations. These results further reveal spin-differentiated behavior in Fe$_{3}$GeTe$_{2}$, where spin-up states are majorly responsible for incoherent-coherent crossover with lowering temperature, whereas spin-down states are already in the coherent regime. Interestingly, finite spin splitting and the ordered moment persists even at temperature larger than 4$T_C$, implies non-Stoner magnetism in Fe$_{3}$GeTe$_{2}$. Additionally, large Sommerfeld coefficient obtained in the low temperature limit from DFT+DMFT calculations are commensurate with results obtained from specific heat measurements resolving prior inconsistencies. 
	
	\section{methodology} 
	
	High-quality single crystals of Fe$_{3}$GeTe$_{2}$ were prepared using chemical vapor transport method with I$_{2}$ as transport agent \cite{chowdhury_unconventional_2021,roy_chowdhury_modification_2022}. Direction dependent magnetic measurements reveal the average Curie temperature, $T_{C}$ to be 206 $\pm$ 4 K \cite{chowdhury_unconventional_2021,roy_chowdhury_modification_2022}. Room temperature  lattice parameters were found to be $a$ = $b$ = 3.99 \AA ~and $c$ = 16.33 \AA ~ in good agreement with previous report \cite{chen_magnetic_2013}. In the photoemission spectroscopic measurements, the Fermi level ($E_{F}$) positions and energy resolutions for various radiations were obtained by measuring the Fermi-edge of a clean polycrystalline silver sample at 30 K. Total energy resolutions were set to 300 meV, 12 meV and 5 meV for Al $K_{\alpha}$ ($h\nu$ = 1486.6 eV), He~{\scriptsize II} ($h\nu$ = 40.8 eV) and  He {\scriptsize I} ($h\nu$ = 21.2 eV) radiations (energy), respectively. Multiple single crystals of Fe$_{3}$GeTe$_{2}$ were cleaved \textit{in-situ} at base pressure better than 4$\times$10$^{-11}$ mbar, to ascertain the cleanliness of the sample surface and reproducibility of the data (see Supplemental Material (SM) \cite{SM} for survey scan and low energy electron diffraction (LEED)). 
	
	Electronic structure calculations were performed using experimental lattice parameters having two formula units (f.u.) per unit cell. Full-potential linearized augmented plane wave method as implemented in \textsc{Wien2k} \cite{Wien2k2020} was used for the DFT calculations. Generalized gradient approximation (GGA) of Purdew-Burke-Ernzerhof \cite{perdew_generalized_1996} was employed for the exchange correlation functional. 18 $\times$ 18 $\times$ 3 $k$-mesh within the first Brillouin zone was used for the self-consistent calculations. The energy and charge convergence criteria were set to $10^{-4}$ eV and $10^{-4}$ electronic charge per f.u., respectively. eDMFT code \cite{haule_dynamical_2010} was used for the charge self-consistent DFT+DMFT calculations, with two impurity problems for Fe I and Fe II, and all five Fe 3$d$ orbitals (forming three non-degenerate groups, $d_{{z}^{2}}$, $d_{{{x}^{2}}-{{y}^{2}}}$/$d_{xy}$ and $d_{xz}$/$d_{yz}$) were considered into correlated sub-space. The Continuous-Time Quantum Monte Carlo (CTQMC) impurity solver \cite{haule_quantum_2007} was used with double counting correction as ‘\textit{exact}’ \cite{haule_exact_2015}. The Hubbard-$U$ = 5.0 eV and Hund's coupling-$J$ = 0.9 eV was opted for both Fe I and Fe II, in accordance with earlier report \cite{kim_large_2018}. Analytical continuation was performed using maximum entropy method \cite{haule_dynamical_2010} to calculate self-energy on the real axis.   
	\begin{figure*}
		\centerline{\includegraphics[width=0.65\textwidth]{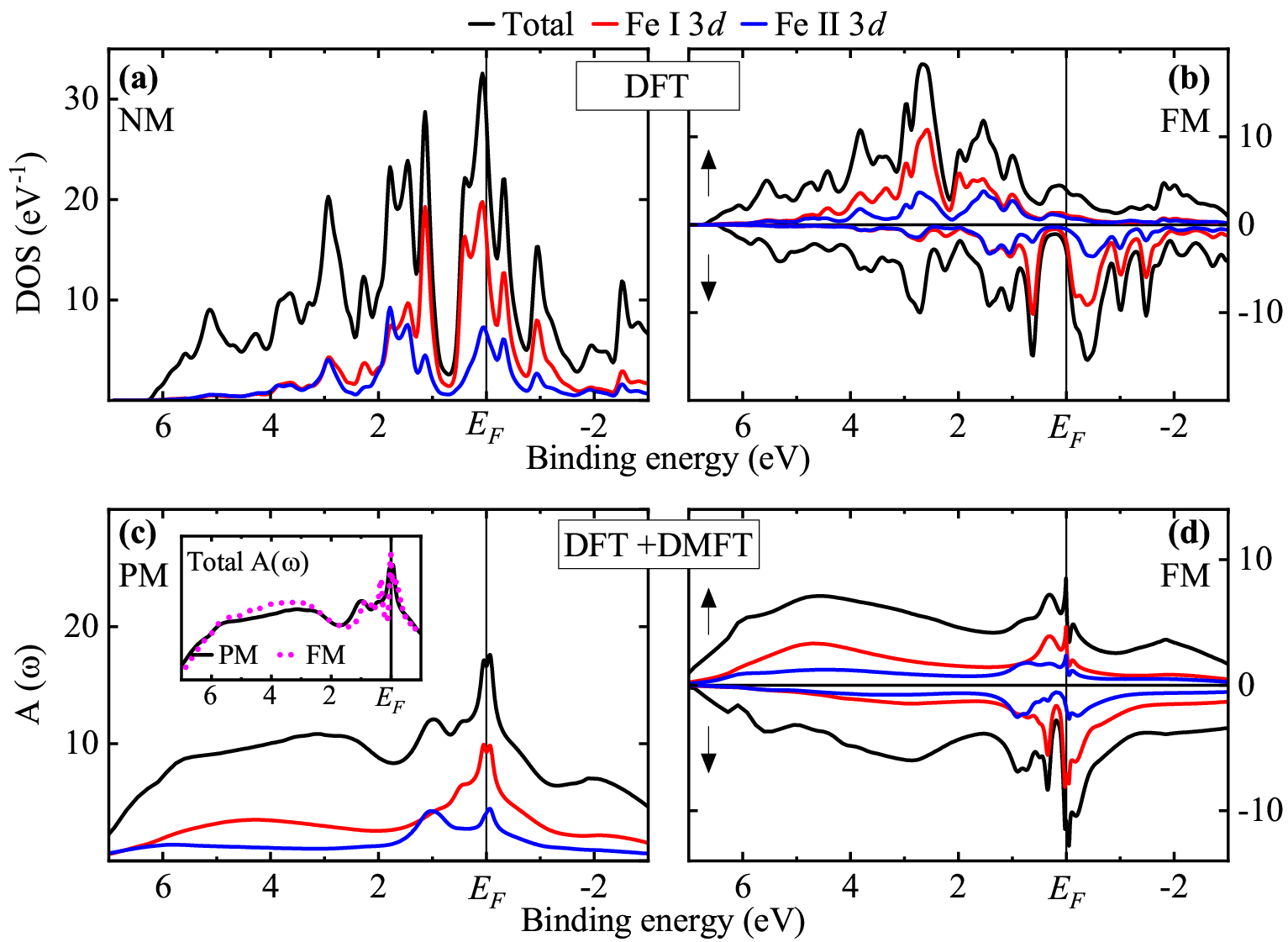}}
		\caption{Total DOS and partial DOSs of Fe I and Fe II for Fe$_{3}$GeTe$_{2}$ using (a) NM and (b) FM DFT. Total and partial spectral functions using DFT+DMFT in (c) PM (\textit{T} = 300 K) and (d) FM (\textit{T} = 50 K) phases. Inset in (c) shows the comparison of total spectral functions obtained in PM and FM DFT+DMFT phases.}\label{fig:fig2}
	\end{figure*}
	
	\section{results and discussions}
	
	The crystal structure and Brillouin zone of Fe$_{3}$GeTe$_{2}$ are shown in Fig. \ref{fig:fig1} (a). Each vdW bonded layer comprises a Fe$_{3}$Ge slab sandwiched between Te layers. Two crystallographically different Fe sites are shown in pink (Fe I ) and golden (Fe II) spheres  \cite{zhao_kondo_2021,chowdhury_unconventional_2021,zhuang_strong_2016}. High-quality of the sample and clean surface obtained by \textit{in-situ} cleaving was ascertained by the sharpness of peaks and the absence of any oxide-related features in the core level spectra, as shown in Fig. \ref{fig:fig1} (b) (also see Note 1 of SM \cite{SM} for further details). All the core level spectra remain very similar to their elemental counterparts \cite{Handbook}. No appreciable change was observed for all the core levels while going from 300 K to 30 K. Notably, the width of Fe 2\textit{p} core level spectra in Fe$_{3}$GeTe$_{2}$ is comparable to that of ferromagnet iron (shown by blue lines) while it has been found to be much smaller ($\sim$ 0.6 eV) in the non-magnetic systems \cite{PhysRevB.55.11488,PhysRevResearch.3.013151}. The full width at half maxima (FWHM) of the asymmetric Fe 2\textit{p}$_{3/2}$ peak was estimated (after subtracting a Shirley type integral background), where the FWHM was obtained to be $\sim$ 1.2 eV and $\sim$ 1.4 eV for Fe$_{3}$GeTe$_{2}$ and Fe metal, respectively. The larger width arises due to finite exchange splitting leading to observed dichroism in the core level spectra in the case of magnetic systems \cite{PhysRevLett.65.492,PhysRevB.55.11488,PhysRevB.57.993}. No change in Fe 2\textit{p} core level spectra in Fe$_{3}$GeTe$_{2}$ suggests that the exchange splitting does not change appreciably across $T_{C}$, indicating that the non-Stoner behavior may be applicable here. The valence band in Fe$_{3}$GeTe$_{2}$ is formed by the hybridization of mainly Fe 3$d$, Ge 4$p$ and Te 5$p$ states, and has been shown in Fig. \ref{fig:fig1} (c) collected using Al $K_{\alpha}$ and He~{\scriptsize II} radiations across the magnetic phase transition. Al~$K_{\alpha}$ spectra at 300 K exhibit three discernible features A, B, and C at around 6 eV, 3 eV and a broad feature below 2 eV binding energy (BE), respectively. Relative intensity enhancement of feature C with respect to features A and B while going from Al $K_{\alpha}$ spectra to He {\scriptsize II} spectra can be understood considering the larger photoionization cross-section of Fe 3$d$ states to that of Te 5$p$ and Ge 4$p$ states at lower photon energies \cite{yeh_atomic_1985}, confirming the dominant contribution of Fe 3$d$ states in feature C. Also, the broad feature C in Al $K_{\alpha}$ spectra is further resolved in the He {\scriptsize II} spectra, presumably due to better energy resolution, exhibiting a peak near $E_{F}$ and a shoulder structure at 1 eV BE. Surprisingly, the overall valence band spectra also do not show appreciable change across the magnetic phase transition which is in sharp contrast with the prototypical itinerant ferromagnet SrRuO$_{3}$ \cite{kim_nature_2015,10.1063/1.2789731}, except for the states in the close vicinity of $E_{F}$, as shown in the inset of Fig. \ref{fig:fig1} (c) where the signature of a quasiparticle peak (marked by down arrow) at 30 K is evident (will be further discussed in the high-resolution spectra).

	To understand the subtle changes in the electronic structure across the magnetic phase transition, we discuss the results of the DFT calculations. Similar to experimental observations, calculated density of states (DOSs) in non-magnetic (NM) phase reveal predominant Fe 3\textit{d} states between $\pm$ 2 eV BE (shown in Fig. \ref{fig:fig2} (a)), with large states at $E_{F}$ indicating metallic character. The states corresponding to Ge 4$p$ and Te 5$p$ primarily appear at higher BE (Fig. S3 \cite{SM} of SM). As expected, the total energy per unit cell reduces by 130 meV in the ferromagnetic (FM) phase and the resulting spin-polarized DOSs are shown in Fig. \ref{fig:fig2} (b). Large redistribution of exchange split spin-polarized states in FM phase compared to NM phase, substantial Fe 3$d$ states appearing even beyond 2 eV BE, along with much reduced DOS($E_F$), are in sharp contrast with the experimental observations. The obtained average magnetic moment of about 2.2 $\mu_B$/Fe is also an overestimation from the value (1.6 $\mu_B$/Fe) obtained from the magnetic measurements \cite{chowdhury_unconventional_2021}. Even the inclusion of Hubbard-\textit{U} in DFT+\textit{U} does not reproduce the experimental results (see Fig. S4 and S5 of SM \cite{SM}). These observations are consistent with earlier studies, concluding that the DFT and DFT+$U$ fails to achieve an agreement with the experimental lattice parameters, magnetic moment and valence band spectra in the magnetically ordered phase \cite{zhu_electronic_2016,ghosh_unraveling_2023}.

	Having an admixture of localized and itinerant $d$ electrons, Fe$_{3}$GeTe$_{2}$ has been understood as correlated electron system with significant role of Hund's-\textit{J} \cite{kim_fe3gete2_2022}. DFT+DMFT has been quite successful in accurately describing such systems with various magnetic transitions \cite{DFT+DMFTarpitapaul,kotliar_electronic_2006}, since it captures the fluctuating moment in the paramagnetic (PM) phase, along with temperature dependent moment in the magnetically ordered phase \cite{kotliar_electronic_2006,zhou_dynamical_2021,kvashnin_dynamical_2022}. In Fig. \ref{fig:fig2} 
	(c), we show the PM DFT+DMFT results for \textit{T} = 300 K (${\beta}$ = 38.68 eV$^{-1}$), where renormalized Fe 3$d$ bands appear between $\pm$ 1.5 eV BE. The total spectral function exhibiting a peak around $E_{F}$ along with a hump at 1 eV BE are in excellent agreement with the He~{\scriptsize II} spectra at 300 K. The local spin moment was found to be 1.72 $\mu_{B}$/Fe in PM DFT+DMFT, calculated using ${\sum}_{i}~2P{_i} |S{_i}$$^{z}|$  (where $P{_i}$ and $|S{_i}$$^{z}|$ represent the probability and absolute spin moment, respectively). The FM DFT+DMFT result for \textit{T} = 50 K (${\beta}$ = 232 eV$^{-1}$) shown in Fig. \ref{fig:fig2} (d) exhibits quasiparticle peak at $E_{F}$ in the exchange split spin-polarized spectral functions corresponding to both the Fe sites. The total spectral function remains largely unchanged while going from PM to FM phase, as shown in the inset of Fig. \ref{fig:fig2} (c). Further, the magnetic moment of 1.51 $\mu_{B}$/Fe in the FM phase is in close agreement with the experimental saturation moment \cite{chowdhury_unconventional_2021} and closely approaches local spin moment obtained in the PM phase. Thus, DFT+DMFT framework successfully determines the electronic structure of Fe$_{3}$GeTe$_{2}$ in both the PM and FM phases.
	
	\begin{figure}[t]
		\centerline{\includegraphics[width=0.5\textwidth]{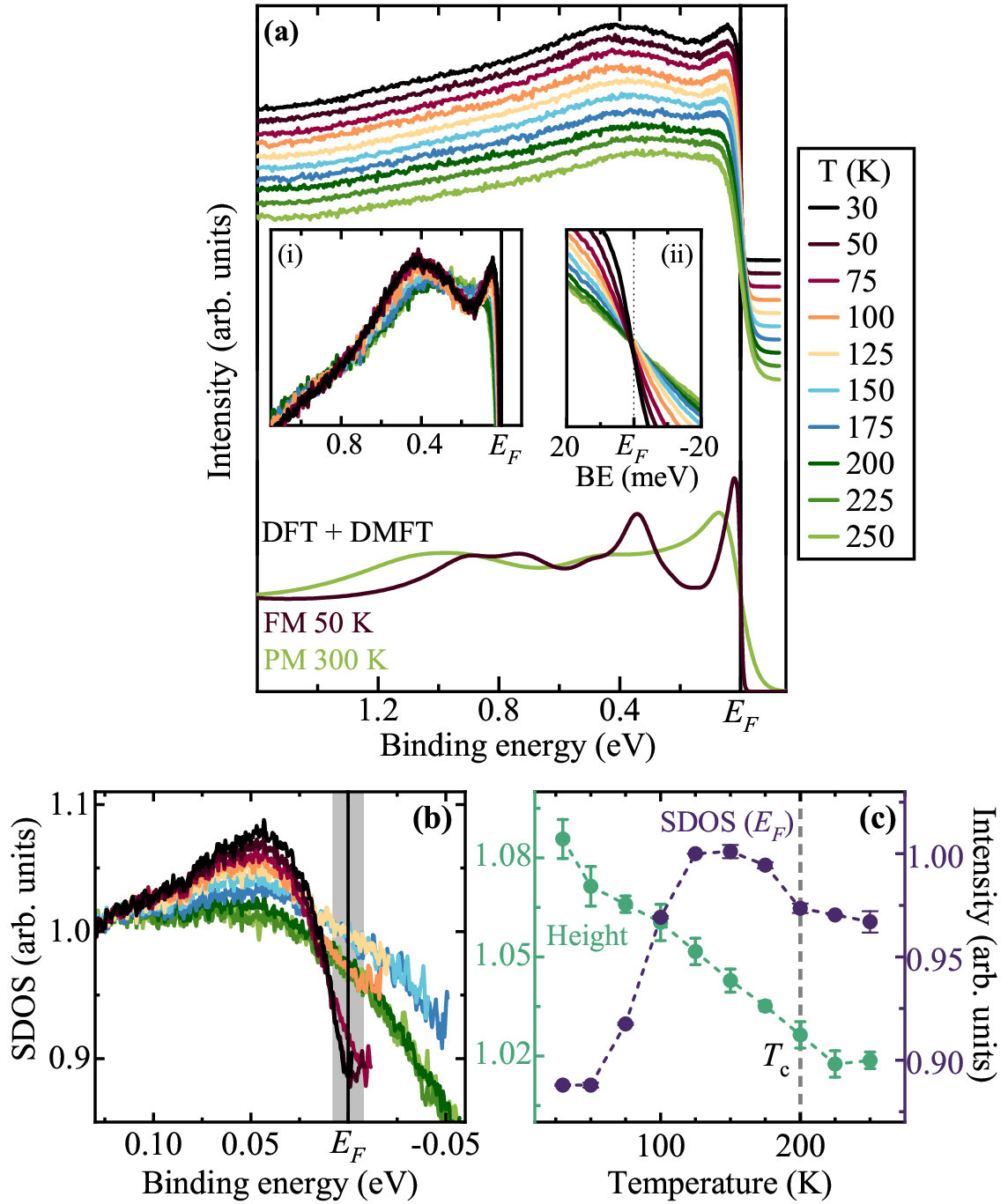}}
		\caption{(a) Temperature dependent high-resolution valence band spectra collected using He {\scriptsize I}. Lower panel shows the occupied part of DFT+DMFT spectral function for PM (300 K) and FM (50 K) phases. Inset (i): \textit{T}-dependent spectral weight redistribution. Inset (ii): \textit{T} evolution of the spectra in the close vicinity of $E_{F}$. \textit{T}-dependent (b) SDOS in the vicinity of $E_{F}$, (c) quasiparticle peak height (green) and SDOS($E_{F}$) (purple) obtained from (b). Shaded region in (b) represents $\pm$ 3$k_BT$ range at 30 K.}\label{fig:fig3}
		\vspace{-2ex}
	\end{figure} 
	
	\begin{figure*}
		\centerline{\includegraphics[width=1.0\textwidth]{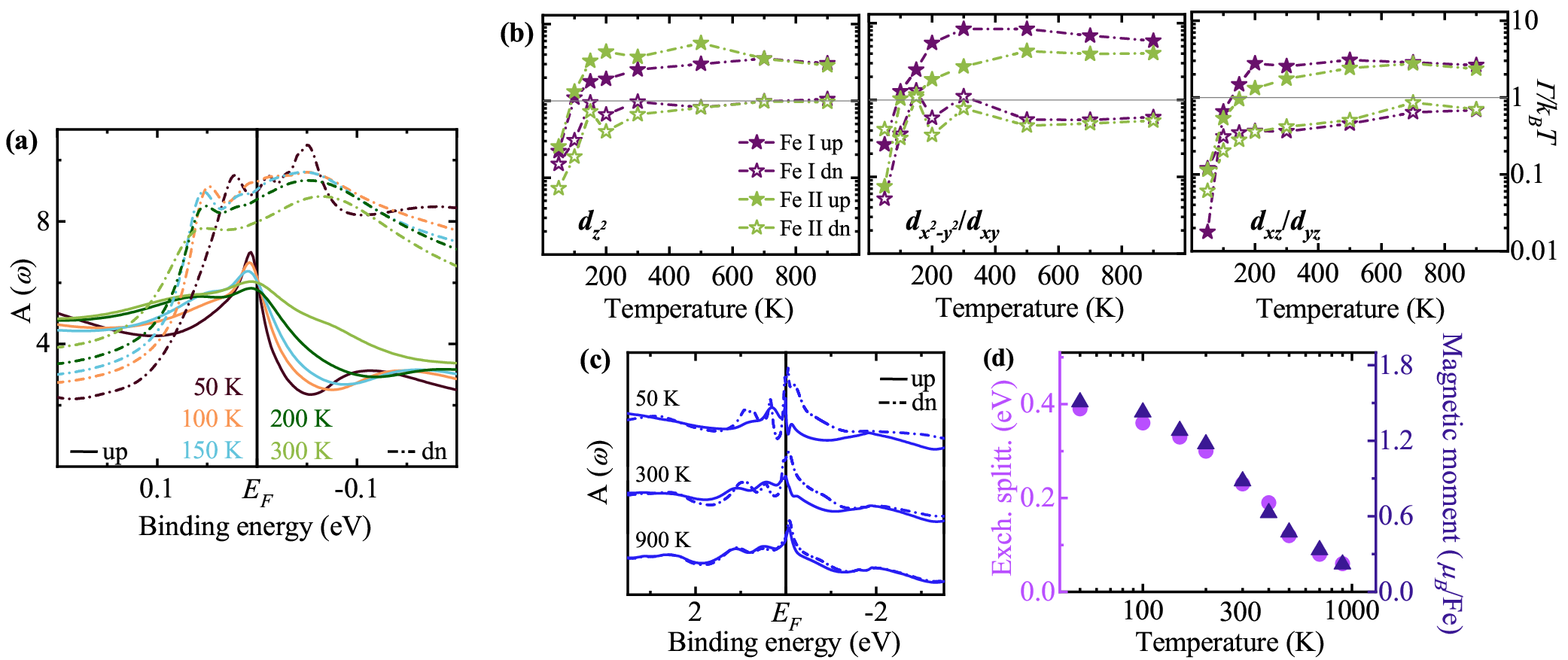}}
		\caption{Temperature dependent FM DFT+DMFT results. (a) Spin-up (solid line) and spin-down (dash-dot line) Fe 3\textit{d} spectral functions. (b) $\varGamma$/\textit{k}$_{B}T$ for the three non-degenerate orbitals (including both the spin) for both the Fe. Spin-up (closed markers) and spin-down (open markers) for Fe I (dark pink) and Fe II (green). (c) Spin-polarized total spectral functions, (d) exchange splitting (circles) and magnetic moments (triangles) at various temperature.}\label{fig:fig4}
	\end{figure*}  
	For further understanding of the evolution of electronic states with temperature, we show the temperature dependent high-resolution valence band spectra collected using He {\scriptsize I} radiation in the top panel of Fig. \ref{fig:fig3} (a). All the spectra are normalized by the total integrated intensity below 1.5 eV BE and have been stacked vertically for clarity. The 250 K spectra exhibit a hump-like structure at $\sim$ 1 eV BE and a broader feature below 0.5 eV BE, similar to high temperature He {\scriptsize II} spectra (Fig. \ref{fig:fig1} (c)). With lowering temperature, the spectral weight redistribution is evident leading to the appearance of sharper quasiparticle peak below 50 meV as shown in the inset (i), while broad feature at 0.4 eV BE remains very similar (within $\pm$ 0.05 eV). It is to be noted here that the energy distribution curves (EDC) at various \textit{k}-points from low temperature ARPES is very similar to the angle-integrated spectra shown here (see APPENDIX I). The lower panel shows the DFT+DMFT spectral function (A($\omega$)) obtained in PM (\textit{T} = 300 K) and FM (\textit{T} = 50 K) phases multiplied by the Fermi-Dirac (FD) function to mimic the occupied states across $T_C$ exhibiting remarkable resemblance of the spectral evolution with the experimental spectra with respect to overall width and energy positions of the features. To visualize the change in the electronic states in the close vicinity of $E_F$, we normalize the spectral intensity at 150 meV and a closer look reveals that the temperature evolution is similar to FD function with a small decrease of the intensity at $E_F$ for low temperature spectra, as shown in the inset (ii). We further show the spectral DOS (SDOS) in Fig. \ref{fig:fig3} (b) obtained by dividing the photoemission intensity with the resolution broadened FD function at respective temperatures \cite{LaCaVO,*BaIrO3,*LSNO}. The evident emergence of quasiparticle peak at $\sim$ 40 meV BE upon entering the FM phase is shown by the monotonous increase in its height and plotted in Fig. \ref{fig:fig3} (c) (using green symbols). Interestingly, SDOS($E_F$) exhibit complex/unusual evolution with temperature, as shown in Fig. \ref{fig:fig3} (c) (using purple symbols). SDOS($E_F$) remains very similar in the PM phase, while it increases upon entering the magnetic phase and achieves a maxima at $\sim$ 125 K, below which it decreases down to 50 K and saturates at lower temperature (also observed in SDOS obtained by symmetrizing the spectra \cite{norman_destruction_1998,*Y2Ir2O7,*reddy_role_2019}, Fig. S2 of SM \cite{SM}). A change of slope in the resistivity of Fe$_{3}$GeTe$_{2}$ near a characteristic temperature (\textit{T}$^{*}$ $ \sim$ 110 K) \cite{zhang_emergence_2018,zhao_kondo_2021} has been associated with an incoherent to coherent crossover similar to the \textit{f}-electron based heavy fermionic systems \cite{hegger_pressure-induced_2000,krellner_relevance_2008,fisk_heavy-electron_1988}. Below $T^*$, the concept of quasiparticle becomes meaningful as the quasiparticle scattering rate, $\varGamma$ (inverse of the lifetime), is smaller than the thermal energy ($k_BT$) and sharper quasiparticle features can be observed \cite{mravlje_coherence-incoherence_2011}. The SDOS obtained from the high-resolution spectra (shown in Fig. \ref{fig:fig3} (b)) reveals decrease in the width of quasiparticle peak below 125 K, also manifested by the sharp decrease of SDOS($E_F$) as shown in Fig. \ref{fig:fig3} (c).

	We additionally showcase and meticulously explore these phenomenon using temperature dependent FM DFT+DMFT calculations. The evolution of the quasiparticle peak with decreasing temperature is also well captured within the FM DFT+DMFT calculations, where the spin-polarized Fe spectral functions are shown in Fig. \ref{fig:fig4} (a). Intriguingly, we observe that the spin-down spectral functions remain largely unchanged around the $E_F$, however, the spin-up spectral functions demonstrate the emergence of quasiparticle peak just below $E_F$ with decreasing temperature, implying significant influence of spin-differentiated electron correlation in Fe$_{3}$GeTe$_{2}$. Further, the scattering rate, $\varGamma$ was obtained for all the orbitals (including spin) for both the Fe sites (see APPENDIX II) and $\varGamma/k_BT$ for various temperature is shown in Fig. \ref{fig:fig4} (b). These results unveiled that the $\varGamma$ for spin-down channels for both the Fe sites remain below $k_BT$, irrespective of temperature, suggesting the coherent scenario, while the spin-up channels show an incoherent-coherent crossover only below 150 K. It is to be noted that, the zero frequency limit of the imaginary part of the self-energy on the imaginary frequency axis, Im$\Sigma$($i\omega\rightarrow 0^+$), and thus the {$\varGamma$} approach to zero faster with decreasing temperature for the spin-up channels than those for the spin-down channels for all the orbitals of both the Fe sites.
	
	Within the itinerant electron magnetism, the exchange splitting of non-degenerate spin bands can be defined via various approaches \cite{kim_nature_2015,kvashnin_exchange_2015,zhuang_strong_2016,xu_signature_2020,ghosh_unraveling_2023}. In the DFT band scenario, the \textit{k}-dependent exchange splitting can be estimated from the difference of the Kohn-Sham eigenvalues of spin split bands and the \textit{k}-averaging gives a reasonable estimate \cite{zhuang_strong_2016}. However, the similar analysis can not be performed in DFT+DMFT due to the diffusive/incoherent nature of the \textit{k}-resolved spectral function \cite{ghosh_unraveling_2023}. Here, we attempt to estimate the exchange splitting in the Fe bands from the energy difference of the centre of the weight of spin-up and spin-down from \textit{k}-integrated spectral functions. The calculated spin-polarized total spectral functions are shown in Fig. \ref{fig:fig4} (c) for different temperatures. The spectral function at 50 K reveal large exchange splitting, which reduces with increasing temperature where, both spin-up and spin-down spectral functions try to acquire similar structure. Interestingly, the spin-dependent spectral functions remain non-degenerate even above $T_C$ and further upto 900 K. The estimated exchange splitting and obtained magnetic moment also follows the similar trend, and remains finite upto 900 K, as shown in Fig. \ref{fig:fig4} (d).
	
	\begin{figure}[t]
		\centerline{\includegraphics[width=0.46\textwidth]{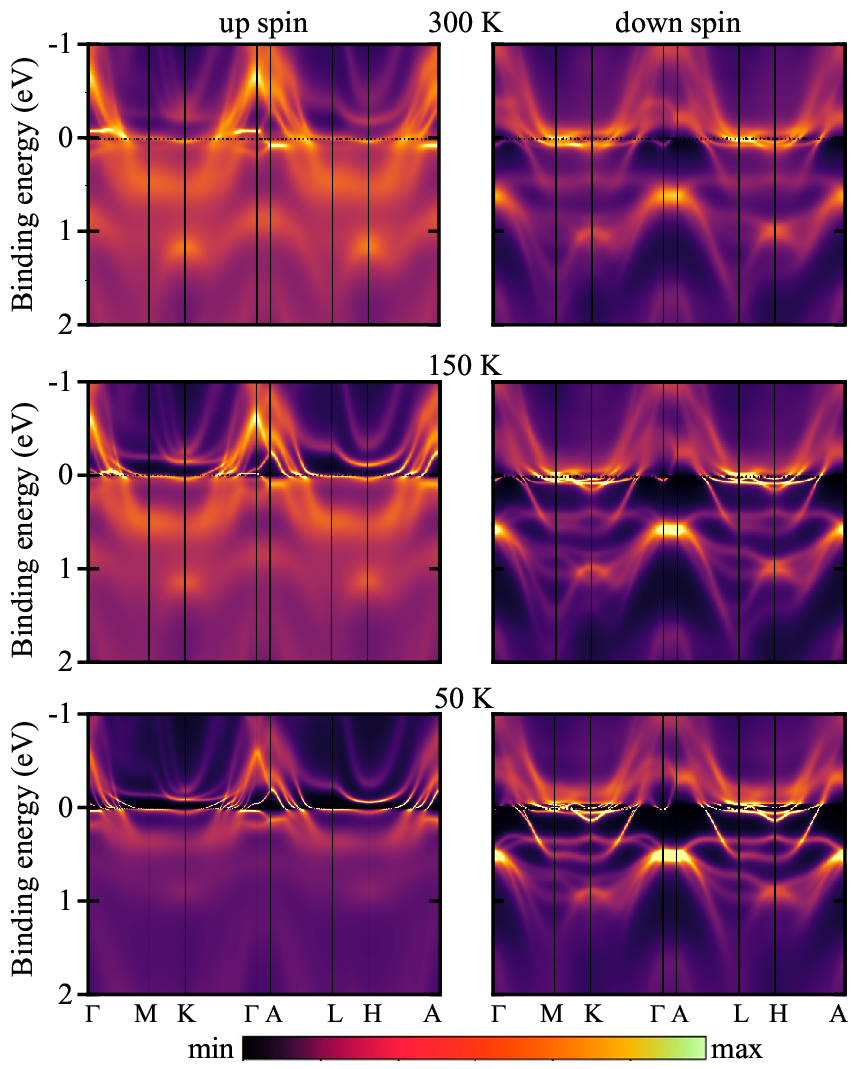}}
		\caption{$k$-resolved spectral functions from FM DFT+DMFT for the spin-up (left) and spin-down (right) at 300 K, 150 K and 50 K.}\label{fig:fig5}
	\end{figure}
	
	To further elucidate the spin-differentiated dual nature, $k$-resolved spin-polarized spectral functions for different temperature obtained from FM DFT+DMFT calculations are shown in Fig. \ref{fig:fig5}. Large number of dispersive bands in both the spin channels show finite exchange splitting and confirm the itinerant character, while the non-dispersive bands in the vicinity of $E_F$ in case of spin-down channel, at high-temperature suggests mixture of localized and itinerant electrons in Fe$_{3}$GeTe$_{2}$. Electronic structure calculations within DFT+DMFT framework hugely overestimates the ferromagnetic transition and reveals no significant change in the valence band across $T_C$, hence, providing additional evidence to conclude Fe$_{3}$GeTe$_{2}$ to be non-Stoner where, the temporal and spatial thermal fluctuation leads to disordered moment (itinerant and local both) thereby destroying the long range magnetic order beyond $T_C$. The Stoner model provides framework for the itinerant magnetic materials, however the inaccurate consideration of spin density fluctuation effects results in many limitations like overestimating \textit{T}$_{C}$ and Curie-Weiss behavior at higher temperature \cite{santiago_itinerant_2017,chen_magnetic_2013}. Thus, these results encourage the need for spin-resolved (AR)PES to be examined for further understanding of the electronic structure of Fe$_{3}$GeTe$_{2}$ across \textit{T}$_{C}$. In addition, diffusive bands in spin-up $k$-resolved spectral functions representing incoherent states show crossover to sharp/coherent states with lowering temperature while spin-down channels remain very similar. The spectral function obtained from low temperature DFT+DMFT are in reasonable agreement with experimental band dispersion along M-$\Gamma$-K directions (shown in APPENDIX I). Further, observed dispersive energy bands along $\Gamma$-A direction in $k$-resolved spectral functions at low temperature implies significant interlayer coupling, thus three-dimensionality of electronic structure in Fe$_{3}$GeTe$_{2}$ \cite{xu_signature_2020}. Lower Fermi velocity of spin split bands in Fe$_{3}$GeTe$_{2}$ than that in the case of a typical metal along with hybridisation of these flat bands with strongly dispersive bands near $E_F$, suggest strengthened quasiparticle mass \cite{rana_spin-polarized_2022,zhang_emergence_2018}.

	\begin{figure}[t]
		\centerline{\includegraphics[width=0.48\textwidth]{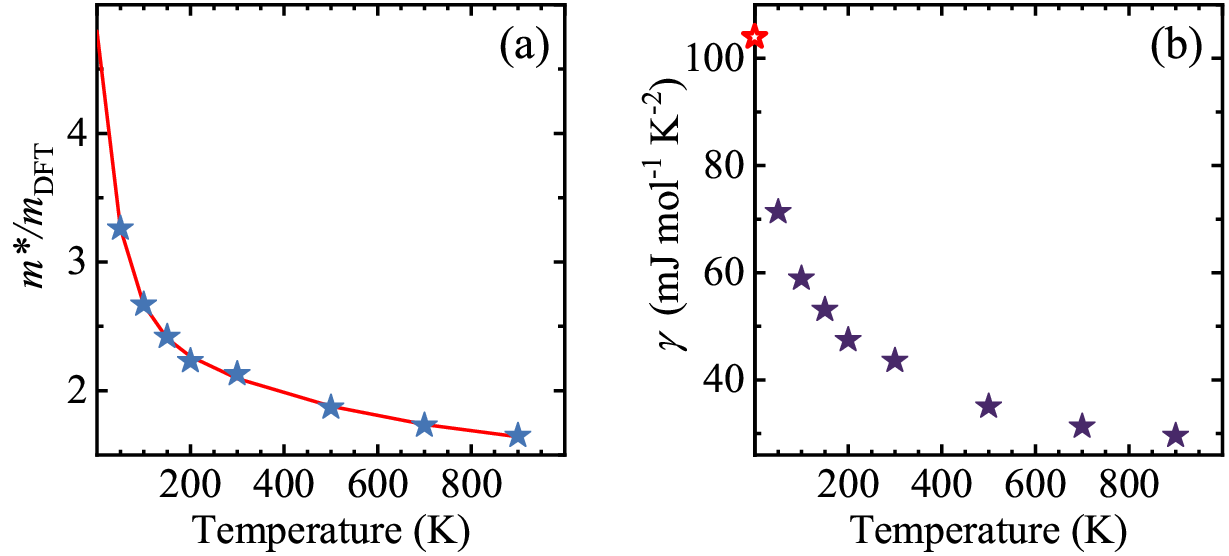}}
		\caption{Calculated (a) mass enhancement factor (${m^*}$/${m_{\text{DFT}}}$) and (b) Sommerfeld coefficient ($\gamma$) using FM DFT+DMFT calculations at various temperature. The red line in (a) shows the fitting using double exponential function (\textit{y}$_{0}$ + \textit{a}$_{1}$$e^{(-x/t1)}$ + \textit{a}$_{2}$$e^{(-x/t2)}$, where \textit{y}$_{0}$=1.45, \textit{a}$_{1}$=2.17, \textit{a}$_{2}$=1.17, \textit{t}1=47.03 and \textit{t}2=498.22) and the open star in (b) shows calculated $\gamma$ obtained using extrapolated value of mass enhancement factor at 0 K.}\label{fig:fig6}
	\end{figure}
	
	Finally, we discuss the discrepancy related to mass enhancement factor (${m^*}$/${m_e}$) in the literature \cite{zhu_electronic_2016,xu_signature_2020,zhang_emergence_2018,kim_large_2018,kim_fe3gete2_2022}. The Sommerfeld coeﬀicient, $\gamma$ (110 mJ mol$^{-1}$ K$^{-2}$) obtained from specific heat measurement is much larger than that of the free electron value of $\approx$ 6 mJ mol$^{-1}$ K$^{-2}$ considering six electron occupancy per Fe, leading to the heavy-fermionic system with ${m^*}$/${m_e}$ of $\approx$ 18 \cite{zhu_electronic_2016,Mizutani}. As discussed earlier, the large reduction of DOS($E_F$) obtained in FM DFT results (\textit{w.r.t.} NM DFT results) is in strong disagreement with PES results and also leads to an overestimation of the $\frac{m^*}{m_{\text{DFT}}}$ = 14.12 (see Fig. S4 of SM \cite{SM}). Within DFT+DMFT framework, $\frac{m^*}{m_{\text{DFT}}}$ is weighted sum of contributions arising from all the orbital ($l$) and spin ($s$), weighted by their local Green’s function ($\propto$ partial DOS) at $E_{F}$ \cite{kotliar_electronic_2006,kim_fe3gete2_2022}. The $\frac{m^*}{m_{\text{DFT}}}$ obtained from FM DFT+DMFT calculations as shown in Fig. \ref{fig:fig6} (a), unveils increasing mass enhancement with lowering temperature and having $\frac{m^*}{m_\text{DFT}}$ = 3.26 at 50 K, suggesting Fe$_{3}$GeTe$_{2}$ to be heavy fermionic system at low temperature. The $\gamma_{_{\text{DMFT}}}$ obtained by linear sum of each contribution, $\gamma_{_{l,s}}$, ($\gamma_{_{\text{DMFT}}}$ = $\sum_{l,s} \gamma_{_{l,s}}$ = $\sum_{l,s}$ [($\pi^2 k_B^2$/3) $\frac{m^*}{m_b}$|$_{_{l,s}}$ PDOS$_{l,s}$ ] $\approx$  [($\pi^2 k_B^2$/3) $\frac{m^*}{m_\text{DFT}}$ A$(\omega)$] \cite{kotliar_electronic_2006} lead to $\sim$ 70 mJ mol$^{-1}$ K$^{-2}$ at 50 K. The extrapolated value of $\frac{m^*}{m_{\text{DFT}}}$ of 4.79 at 0 K leads to $\gamma$ of about $\sim$ 104 mJ mol$^{-1}$ K$^{-2}$, as shown in Fig. \ref{fig:fig6} (b) and is in close agreement with experiments, finally resolving the much debated disparity \cite{zhu_electronic_2016}.
	
	\section{conclusion}
	In summary, electronic structure of Fe$_{3}$GeTe$_{2}$ has been investigated using photoemission spectroscopy and theoretical calculations within DFT, DFT+\textit{U} and DFT+DMFT frameworks. The high-resolution valence band spectra are well captured within DFT+DMFT across the magnetic phase transition. Temperature dependent high-resolution spectra exhibit emergence of quasiparticle peak in close vicinity of $E_{F}$ along with the manifestation of incoherent-coherent crossover where, an anomalous behaviour of spectral density of states at $E_{F}$ is observed $\sim$ 125 K. DFT+DMFT successfully demonstrate the evolution of spin bands with incoherent-coherent crossover along with the increasing effective mass at lower temperatures, concluding the heavy-fermionic nature. We also resolves the long standing issue of large Sommerfeld coefficient in this system obtained within DFT+DMFT calculations. In particular, (i) no significant change in the experimental Fe 2\textit{p} core level and overall valence band across $T_C$, and (ii) finite exchange splitting (also magnetic moment) persisting even beyond 4$T_{C}$ in temperature dependent ferromagnetic DFT+DMFT calculations, together suggests Fe$_{3}$GeTe$_{2}$ to be a non-Stoner ferromagnet. Results presented here advances the understanding of complex evolution of electronic structure and non-Stoner magnetic behavior and lays the foundation for further spin-resolved PES in this correlated van der Waals ferromagnet Fe$_{3}$GeTe$_{2}$.

	\section*{acknowledgments}
	
	D. S. and N. B. acknowledge the Council of Scientific and Industrial Research (CSIR), Government of India, for financial support with Award No. 09/1020(0198)/2020-EMR-I and 09/1020(0177)/2019-EMR-I, respectively. R.R.C. acknowledges the Department of Science and Technology (DST), Government of India, for financial support (Grant No. DST/INSPIRE/04/2018/001755). R.P.S. acknowledges the Science and Engineering Research Board (SERB), Government of India, for Core Research Grant No. CRG/2019/001028. We gratefully acknowledge the use of HPC facility and CIF at IISER Bhopal.

	%\section*{APPENDIX I}
	\section*{\label{sec:level1} APPENDIX I}
	
	\begin{figure*}[h]
		\centerline{\includegraphics[width=0.85\textwidth]{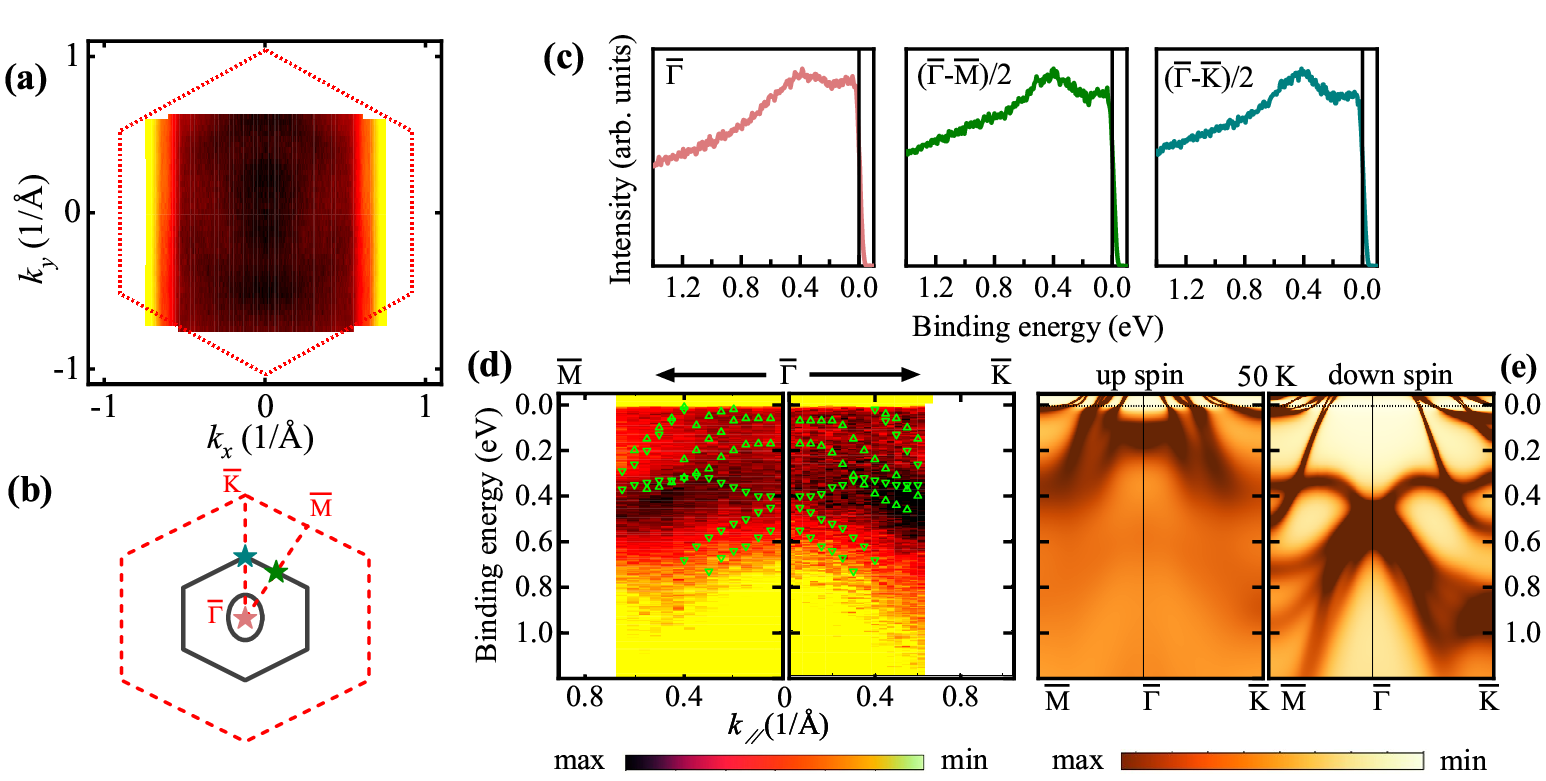}}
		\caption{Fermi surface map of Fe$_{3}$GeTe$_{2}$ (a) using He {\scriptsize I} radiation at 30 K, and (b) schematic diagram. The red dashed line shows the Brillouin zone with high-symmetry directions in (b). The stars marked in (b) with pink, green and blue colour represents $\overline{\mbox{$\Gamma$}}$, ($\overline{\mbox{$\Gamma$}}$-$\overline{\mbox{M}}$)/2 and ($\overline{\mbox{$\Gamma$}}$-$\overline{\mbox{K}}$)/2 locations, respectively, along which (c) energy distribution curves are shown. (d) ARPES band dispersion and (e) \textit{k}-resolved spectral function at $k_z$ = 0.353c$^{*}$ from FM DFT+DMFT (\textit{T} = 50 K) for both the spin bands, along the high-symmetry directions. For better visualization the high intensity bands from up and down spin of \textit{k}-resolved spectral functions, were overlapped on ARPES band dispersion using up and down green triangles, respectively.}\label{fig:fig7}
	\end{figure*}
	
	ARPES Fermi surface map (integrated within $E_F$ $\pm$ 10 meV energy window and symmetrized along $k_x$ = 0 line) obtained using He {\scriptsize I} radiation ($k_z$ = 0.353c$^{*}$, considering inner potential, V$_0$=13.5 eV \cite{xu_signature_2020}) at 30 K (with total energy resolution $\sim$ 15 meV) is shown in Fig. \ref{fig:fig7} (a). The circular and hexagonal shaped Fermi surfaces centered around  $\overline{\mbox{$\Gamma$}}$ is in agreement with earlier observations \cite{xu_signature_2020,zhang_emergence_2018}. Fig. \ref{fig:fig7}(b) shows the schematic of the obtained Fermi surfaces where the dotted red hexagon represents the surface Brillouin zone with high symmetry lines $\overline{\mbox{$\Gamma$}}$-$\overline{\mbox{K}}$, and $\overline{\mbox{$\Gamma$}}$-$\overline{\mbox{M}}$. The stars represents $\overline{\mbox{$\Gamma$}}$, ($\overline{\mbox{$\Gamma$}}$-$\overline{\mbox{M}}$)/2 and ($\overline{\mbox{$\Gamma$}}$-$\overline{\mbox{K}}$)/2, corresponding to which the energy distributive curves (EDCs) (integrated for $\Delta k_x$ and $\Delta k_y$ within $\pm$ 0.02 \AA $^{-1}$) are shown in Fig. \ref{fig:fig7} (c). Strengthened quasiparticle peak ($\sim$ 40 meV BE) is evident from the EDC at $\overline{\mbox{$\Gamma$}}$, as compared to the same obtained from EDC at ($\overline{\mbox{$\Gamma$}}$-$\overline{\mbox{M}}$)/2 and ($\overline{\mbox{$\Gamma$}}$-$\overline{\mbox{K}}$)/2. It is to be noted that the overall EDCs at various locations across Brillouin zone are mostly similar and agrees well with the momentum-integration high-resolution PES spectra as presented in the main text (Fig. \ref{fig:fig3} (a)). Further, Fig. \ref{fig:fig7} (d) and (e) represent the band dispersion from ARPES (at 30 K) and \textit{k}-resolved spectral functions of both the spins from FM DFT+DMFT calculation (at 50 K, $k_z$ = 0.353c$^{*}$), respectively, along high-symmetry directions. ARPES spectra (Fig. \ref{fig:fig7} (d)) shows broad (incoherent like) bands at around 0.4 eV BE and sharper (coherent like) bands around the $E_{F}$ in $\overline{\mbox{M}}$-$\overline{\mbox{$\Gamma$}}$ as well as in $\overline{\mbox{$\Gamma$}}$-$\overline{\mbox{K}}$ directions, which are better resolved in both the spin channels of \textit{k}-resolved spectral functions of DFT+DMFT. For better visualization, the bands with higher spectral functions from DFT+DMFT results were overlapped on the ARPES spectra (using markers) and is shown in Fig. \ref{fig:fig7} (d) with up and down green triangles for up spin and for down spin, respectively.
	
	\section*{\label{sec:level2} APPENDIX II}
	
	\begin{figure*}[h]
		\centerline{\includegraphics[width=0.6\textwidth]{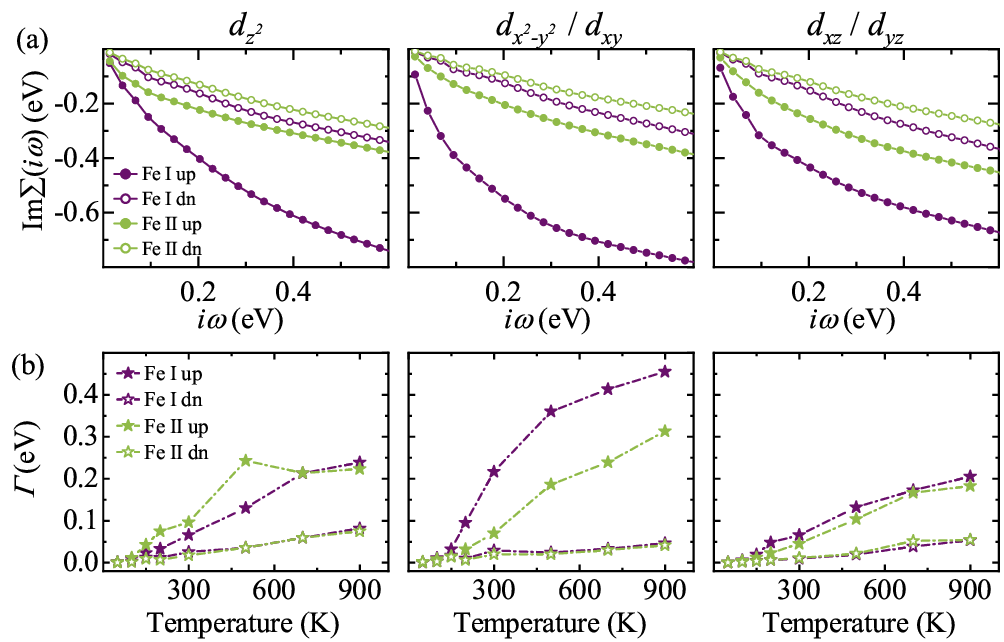}}
		\caption{(a) Spin and orbital dependent imaginary part of self-energy on Matsubara frequency axis, Im${\Sigma}$($i{\omega}$) for both the Fe using FM DFT+DMFT calculations at 50 K. (b) Quasiparticle scattering rate ($\varGamma$) for the three non-degenerate orbitals (including spin) for both the Fe at various temperatures. Spin-up and spin-down are shown in closed markers and open markers, respectively for Fe I (dark pink) and Fe II (green).}\label{fig:fig8}
	\end{figure*}
	
	Within FM DFT+DMFT calculation the spin and orbital dependent imaginary part of the self-energy on the Matsubara frequency axis, Im${\Sigma}$($i{\omega}$) for both Fe sites were obtained and are shown in Fig. \ref{fig:fig8} (a) for \textit{T} = 50 K. The Im${\Sigma}$($i{\omega}$) approaches to zero at lower frequency for all the orbitals and spins of both the Fe. The quasiparticle scattering rate is calculated using $\varGamma$$_{l,s}$ = $-(m^{*}$/$m_{_{b}}$)$_{l,s}$ $^{-1}$ Im${\Sigma}$$_{l,s}$ ($i{\omega}$${\rightarrow 0^{+}}$), where ( $m^{*}$/$m_{_{b}}$)$_{l,s}$ = 1 $-~{\partial}$ Im${\Sigma}$$_{l,s}$($i{\omega}$)/${\partial}$(${i\omega}$)$|_{i\omega}$$_{\rightarrow 0}$ for $l$ orbitals, and \textit{s} spins, and is shown in Fig. \ref{fig:fig8} (b) \cite{kotliar_electronic_2006,kim_fe3gete2_2022} for various temperatures. The imaginary part of self energy at the zero frequency limit and its derivative was obtained using fourth order polynomial fit for the first six data points. The spin-up channels for both the Fe exhibit larger Im${\Sigma}$$_{l}$ ($i{\omega}$${\rightarrow 0^{+}}$) for all the orbitals at higher temperature and tends to zero faster at lower temperature in comparison to that for the spin down channels, implying significant influence of spin-differentiated electron correlation in Fe$_{3}$GeTe$_{2}$. An overall trend of $\varGamma$$_{l,s}$ is consistent with the imaginary part of self energy at the zero frequency limit. Further, the $\varGamma$ obtained for all the orbitals for both the Fe sites, unveiled that the spin-down channels remain below $k_BT$ (as shown in Fig. \ref{fig:fig4} (b)), irrespective of temperature for both the Fe sites, suggesting the coherent scenario throughout, while the spin-up channels show an incoherent-coherent crossover only below 150 K.
	
	%\bibliography{BIB.bib}

\begin{thebibliography}{59}%
		\makeatletter
		\providecommand \@ifxundefined [1]{%
			\@ifx{#1\undefined}
		}%
		\providecommand \@ifnum [1]{%
			\ifnum #1\expandafter \@firstoftwo
			\else \expandafter \@secondoftwo
			\fi
		}%
		\providecommand \@ifx [1]{%
			\ifx #1\expandafter \@firstoftwo
			\else \expandafter \@secondoftwo
			\fi
		}%
		\providecommand \natexlab [1]{#1}%
		\providecommand \enquote  [1]{``#1''}%
		\providecommand \bibnamefont  [1]{#1}%
		\providecommand \bibfnamefont [1]{#1}%
		\providecommand \citenamefont [1]{#1}%
		\providecommand \href@noop [0]{\@secondoftwo}%
		\providecommand \href [0]{\begingroup \@sanitize@url \@href}%
		\providecommand \@href[1]{\@@startlink{#1}\@@href}%
		\providecommand \@@href[1]{\endgroup#1\@@endlink}%
		\providecommand \@sanitize@url [0]{\catcode `\\12\catcode `\$12\catcode
			`\&12\catcode `\#12\catcode `\^12\catcode `\_12\catcode `\%12\relax}%
		\providecommand \@@startlink[1]{}%
		\providecommand \@@endlink[0]{}%
		\providecommand \url  [0]{\begingroup\@sanitize@url \@url }%
		\providecommand \@url [1]{\endgroup\@href {#1}{\urlprefix }}%
		\providecommand \urlprefix  [0]{URL }%
		\providecommand \Eprint [0]{\href }%
		\providecommand \doibase [0]{https://doi.org/}%
		\providecommand \selectlanguage [0]{\@gobble}%
		\providecommand \bibinfo  [0]{\@secondoftwo}%
		\providecommand \bibfield  [0]{\@secondoftwo}%
		\providecommand \translation [1]{[#1]}%
		\providecommand \BibitemOpen [0]{}%
		\providecommand \bibitemStop [0]{}%
		\providecommand \bibitemNoStop [0]{.\EOS\space}%
		\providecommand \EOS [0]{\spacefactor3000\relax}%
		\providecommand \BibitemShut  [1]{\csname bibitem#1\endcsname}%
		\let\auto@bib@innerbib\@empty
		%</preamble>
		\bibitem [{\citenamefont {Mak}\ \emph {et~al.}(2019)\citenamefont {Mak},
			\citenamefont {Shan},\ and\ \citenamefont {Ralph}}]{mak_probing_2019}%
		\BibitemOpen
		\bibfield  {author} {\bibinfo {author} {\bibfnamefont {K.~F.}\ \bibnamefont
				{Mak}}, \bibinfo {author} {\bibfnamefont {J.}~\bibnamefont {Shan}},\ and\
			\bibinfo {author} {\bibfnamefont {D.~C.}\ \bibnamefont {Ralph}},\ }\bibfield
		{title} {\bibinfo {title} {Probing and controlling magnetic states in {2D}
				layered magnetic materials},\ }\href
		{https://www.nature.com/articles/s42254-019-0110-y} {\bibfield  {journal}
			{\bibinfo  {journal} {Nature Reviews Physics}\ }\textbf {\bibinfo {volume}
				{1}},\ \bibinfo {pages} {646} (\bibinfo {year} {2019})}\BibitemShut {NoStop}%
		\bibitem [{\citenamefont {Gibertini}\ \emph {et~al.}(2019)\citenamefont
			{Gibertini}, \citenamefont {Koperski}, \citenamefont {Morpurgo},\ and\
			\citenamefont {Novoselov}}]{gibertini_magnetic_2019}%
		\BibitemOpen
		\bibfield  {author} {\bibinfo {author} {\bibfnamefont {M.}~\bibnamefont
				{Gibertini}}, \bibinfo {author} {\bibfnamefont {M.}~\bibnamefont {Koperski}},
			\bibinfo {author} {\bibfnamefont {A.~F.}\ \bibnamefont {Morpurgo}},\ and\
			\bibinfo {author} {\bibfnamefont {K.~S.}\ \bibnamefont {Novoselov}},\
		}\bibfield  {title} {\bibinfo {title} {Magnetic {2D} materials and
				heterostructures},\ }\href
		{https://www.nature.com/articles/s41565-019-0438-6} {\bibfield  {journal}
			{\bibinfo  {journal} {Nature Nanotechnology}\ }\textbf {\bibinfo {volume}
				{14}},\ \bibinfo {pages} {408} (\bibinfo {year} {2019})}\BibitemShut
		{NoStop}%
		\bibitem [{\citenamefont {Burch}\ \emph {et~al.}(2018)\citenamefont {Burch},
			\citenamefont {Mandrus},\ and\ \citenamefont {Park}}]{burch_magnetism_2018}%
		\BibitemOpen
		\bibfield  {author} {\bibinfo {author} {\bibfnamefont {K.~S.}\ \bibnamefont
				{Burch}}, \bibinfo {author} {\bibfnamefont {D.}~\bibnamefont {Mandrus}},\
			and\ \bibinfo {author} {\bibfnamefont {J.-G.}\ \bibnamefont {Park}},\
		}\bibfield  {title} {\bibinfo {title} {Magnetism in two-dimensional van der
				{Waals} materials},\ }\href
		{https://www.nature.com/articles/s41586-018-0631-z} {\bibfield  {journal}
			{\bibinfo  {journal} {Nature}\ }\textbf {\bibinfo {volume} {563}},\ \bibinfo
			{pages} {47} (\bibinfo {year} {2018})}\BibitemShut {NoStop}%
		\bibitem [{\citenamefont {Gong}\ \emph {et~al.}(2017)\citenamefont {Gong},
			\citenamefont {Li}, \citenamefont {Li}, \citenamefont {Ji}, \citenamefont
			{Stern}, \citenamefont {Xia}, \citenamefont {Cao}, \citenamefont {Bao},
			\citenamefont {Wang}, \citenamefont {Wang}, \citenamefont {Qiu},
			\citenamefont {Cava}, \citenamefont {Louie}, \citenamefont {Xia},\ and\
			\citenamefont {Zhang}}]{gong_discovery_2017}%
		\BibitemOpen
		\bibfield  {author} {\bibinfo {author} {\bibfnamefont {C.}~\bibnamefont
				{Gong}}, \bibinfo {author} {\bibfnamefont {L.}~\bibnamefont {Li}}, \bibinfo
			{author} {\bibfnamefont {Z.}~\bibnamefont {Li}}, \bibinfo {author}
			{\bibfnamefont {H.}~\bibnamefont {Ji}}, \bibinfo {author} {\bibfnamefont
				{A.}~\bibnamefont {Stern}}, \bibinfo {author} {\bibfnamefont
				{Y.}~\bibnamefont {Xia}}, \bibinfo {author} {\bibfnamefont {T.}~\bibnamefont
				{Cao}}, \bibinfo {author} {\bibfnamefont {W.}~\bibnamefont {Bao}}, \bibinfo
			{author} {\bibfnamefont {C.}~\bibnamefont {Wang}}, \bibinfo {author}
			{\bibfnamefont {Y.}~\bibnamefont {Wang}}, \bibinfo {author} {\bibfnamefont
				{Z.~Q.}\ \bibnamefont {Qiu}}, \bibinfo {author} {\bibfnamefont {R.~J.}\
				\bibnamefont {Cava}}, \bibinfo {author} {\bibfnamefont {S.~G.}\ \bibnamefont
				{Louie}}, \bibinfo {author} {\bibfnamefont {J.}~\bibnamefont {Xia}},\ and\
			\bibinfo {author} {\bibfnamefont {X.}~\bibnamefont {Zhang}},\ }\bibfield
		{title} {\bibinfo {title} {Discovery of intrinsic ferromagnetism in
				two-dimensional van der {Waals} crystals},\ }\href
		{https://www.nature.com/articles/nature22060} {\bibfield  {journal} {\bibinfo
				{journal} {Nature}\ }\textbf {\bibinfo {volume} {546}},\ \bibinfo {pages}
			{265} (\bibinfo {year} {2017})}\BibitemShut {NoStop}%
		\bibitem [{\citenamefont {Williams}\ \emph {et~al.}(2015)\citenamefont
			{Williams}, \citenamefont {Aczel}, \citenamefont {Lumsden}, \citenamefont
			{Nagler}, \citenamefont {Stone}, \citenamefont {Yan},\ and\ \citenamefont
			{Mandrus}}]{williams_magnetic_2015}%
		\BibitemOpen
		\bibfield  {author} {\bibinfo {author} {\bibfnamefont {T.~J.}\ \bibnamefont
				{Williams}}, \bibinfo {author} {\bibfnamefont {A.~A.}\ \bibnamefont {Aczel}},
			\bibinfo {author} {\bibfnamefont {M.~D.}\ \bibnamefont {Lumsden}}, \bibinfo
			{author} {\bibfnamefont {S.~E.}\ \bibnamefont {Nagler}}, \bibinfo {author}
			{\bibfnamefont {M.~B.}\ \bibnamefont {Stone}}, \bibinfo {author}
			{\bibfnamefont {J.-Q.}\ \bibnamefont {Yan}},\ and\ \bibinfo {author}
			{\bibfnamefont {D.}~\bibnamefont {Mandrus}},\ }\bibfield  {title} {\bibinfo
			{title} {Magnetic correlations in the quasi-two-dimensional semiconducting
				ferromagnet {CrSiTe$_3$}},\ }\href
		{https://link.aps.org/doi/10.1103/PhysRevB.92.144404} {\bibfield  {journal}
			{\bibinfo  {journal} {Phys. Rev. B}\ }\textbf {\bibinfo {volume} {92}},\
			\bibinfo {pages} {144404} (\bibinfo {year} {2015})}\BibitemShut {NoStop}%
		\bibitem [{\citenamefont {Huang}\ \emph {et~al.}(2017)\citenamefont {Huang},
			\citenamefont {Clark}, \citenamefont {Navarro-Moratalla}, \citenamefont
			{Klein}, \citenamefont {Cheng}, \citenamefont {Seyler}, \citenamefont
			{Zhong}, \citenamefont {Schmidgall}, \citenamefont {McGuire}, \citenamefont
			{Cobden}, \citenamefont {Yao}, \citenamefont {Xiao}, \citenamefont
			{Jarillo-Herrero},\ and\ \citenamefont {Xu}}]{huang_layer-dependent_2017}%
		\BibitemOpen
		\bibfield  {author} {\bibinfo {author} {\bibfnamefont {B.}~\bibnamefont
				{Huang}}, \bibinfo {author} {\bibfnamefont {G.}~\bibnamefont {Clark}},
			\bibinfo {author} {\bibfnamefont {E.}~\bibnamefont {Navarro-Moratalla}},
			\bibinfo {author} {\bibfnamefont {D.~R.}\ \bibnamefont {Klein}}, \bibinfo
			{author} {\bibfnamefont {R.}~\bibnamefont {Cheng}}, \bibinfo {author}
			{\bibfnamefont {K.~L.}\ \bibnamefont {Seyler}}, \bibinfo {author}
			{\bibfnamefont {D.}~\bibnamefont {Zhong}}, \bibinfo {author} {\bibfnamefont
				{E.}~\bibnamefont {Schmidgall}}, \bibinfo {author} {\bibfnamefont {M.~A.}\
				\bibnamefont {McGuire}}, \bibinfo {author} {\bibfnamefont {D.~H.}\
				\bibnamefont {Cobden}}, \bibinfo {author} {\bibfnamefont {W.}~\bibnamefont
				{Yao}}, \bibinfo {author} {\bibfnamefont {D.}~\bibnamefont {Xiao}}, \bibinfo
			{author} {\bibfnamefont {P.}~\bibnamefont {Jarillo-Herrero}},\ and\ \bibinfo
			{author} {\bibfnamefont {X.}~\bibnamefont {Xu}},\ }\bibfield  {title}
		{\bibinfo {title} {Layer-dependent ferromagnetism in a van der {Waals}
				crystal down to the monolayer limit},\ }\href
		{https://www.nature.com/articles/nature22391} {\bibfield  {journal} {\bibinfo
				{journal} {Nature}\ }\textbf {\bibinfo {volume} {546}},\ \bibinfo {pages}
			{270} (\bibinfo {year} {2017})}\BibitemShut {NoStop}%
		\bibitem [{\citenamefont {Deng}\ \emph {et~al.}(2018)\citenamefont {Deng},
			\citenamefont {Yu}, \citenamefont {Song}, \citenamefont {Zhang},
			\citenamefont {Wang}, \citenamefont {Sun}, \citenamefont {Yi}, \citenamefont
			{Wu}, \citenamefont {Wu}, \citenamefont {Zhu}, \citenamefont {Wang},
			\citenamefont {Chen},\ and\ \citenamefont {Zhang}}]{deng_gate-tunable_2018}%
		\BibitemOpen
		\bibfield  {author} {\bibinfo {author} {\bibfnamefont {Y.}~\bibnamefont
				{Deng}}, \bibinfo {author} {\bibfnamefont {Y.}~\bibnamefont {Yu}}, \bibinfo
			{author} {\bibfnamefont {Y.}~\bibnamefont {Song}}, \bibinfo {author}
			{\bibfnamefont {J.}~\bibnamefont {Zhang}}, \bibinfo {author} {\bibfnamefont
				{N.~Z.}\ \bibnamefont {Wang}}, \bibinfo {author} {\bibfnamefont
				{Z.}~\bibnamefont {Sun}}, \bibinfo {author} {\bibfnamefont {Y.}~\bibnamefont
				{Yi}}, \bibinfo {author} {\bibfnamefont {Y.~Z.}\ \bibnamefont {Wu}}, \bibinfo
			{author} {\bibfnamefont {S.}~\bibnamefont {Wu}}, \bibinfo {author}
			{\bibfnamefont {J.}~\bibnamefont {Zhu}}, \bibinfo {author} {\bibfnamefont
				{J.}~\bibnamefont {Wang}}, \bibinfo {author} {\bibfnamefont {X.~H.}\
				\bibnamefont {Chen}},\ and\ \bibinfo {author} {\bibfnamefont
				{Y.}~\bibnamefont {Zhang}},\ }\bibfield  {title} {\bibinfo {title}
			{Gate-tunable room-temperature ferromagnetism in two-dimensional
				{Fe$_3$GeTe$_2$}},\ }\href
		{https://www.nature.com/articles/s41586-018-0626-9} {\bibfield  {journal}
			{\bibinfo  {journal} {Nature}\ }\textbf {\bibinfo {volume} {563}},\ \bibinfo
			{pages} {94} (\bibinfo {year} {2018})}\BibitemShut {NoStop}%
		\bibitem [{\citenamefont {Birch}\ \emph {et~al.}(2022)\citenamefont {Birch},
			\citenamefont {Powalla}, \citenamefont {Wintz}, \citenamefont {Hovorka},
			\citenamefont {Litzius}, \citenamefont {Loudon}, \citenamefont {Turnbull},
			\citenamefont {Nehruji}, \citenamefont {Son}, \citenamefont {Bubeck},
			\citenamefont {Rauch}, \citenamefont {Weigand}, \citenamefont {Goering},
			\citenamefont {Burghard},\ and\ \citenamefont
			{Schütz}}]{birch_history-dependent_2022-2}%
		\BibitemOpen
		\bibfield  {author} {\bibinfo {author} {\bibfnamefont {M.~T.}\ \bibnamefont
				{Birch}}, \bibinfo {author} {\bibfnamefont {L.}~\bibnamefont {Powalla}},
			\bibinfo {author} {\bibfnamefont {S.}~\bibnamefont {Wintz}}, \bibinfo
			{author} {\bibfnamefont {O.}~\bibnamefont {Hovorka}}, \bibinfo {author}
			{\bibfnamefont {K.}~\bibnamefont {Litzius}}, \bibinfo {author} {\bibfnamefont
				{J.~C.}\ \bibnamefont {Loudon}}, \bibinfo {author} {\bibfnamefont {L.~A.}\
				\bibnamefont {Turnbull}}, \bibinfo {author} {\bibfnamefont {V.}~\bibnamefont
				{Nehruji}}, \bibinfo {author} {\bibfnamefont {K.}~\bibnamefont {Son}},
			\bibinfo {author} {\bibfnamefont {C.}~\bibnamefont {Bubeck}}, \bibinfo
			{author} {\bibfnamefont {T.~G.}\ \bibnamefont {Rauch}}, \bibinfo {author}
			{\bibfnamefont {M.}~\bibnamefont {Weigand}}, \bibinfo {author} {\bibfnamefont
				{E.}~\bibnamefont {Goering}}, \bibinfo {author} {\bibfnamefont
				{M.}~\bibnamefont {Burghard}},\ and\ \bibinfo {author} {\bibfnamefont
				{G.}~\bibnamefont {Schütz}},\ }\bibfield  {title} {\bibinfo {title}
			{History-dependent domain and skyrmion formation in {2D} van der {Waals}
				magnet {Fe$_3$GeTe$_2$}},\ }\href
		{https://www.nature.com/articles/s41467-022-30740-7} {\bibfield  {journal}
			{\bibinfo  {journal} {Nature Communications}\ }\textbf {\bibinfo {volume}
				{13}},\ \bibinfo {pages} {3035} (\bibinfo {year} {2022})}\BibitemShut
		{NoStop}%
		\bibitem [{\citenamefont {Chowdhury}\ \emph {et~al.}(2021)\citenamefont
			{Chowdhury}, \citenamefont {DuttaGupta}, \citenamefont {Patra}, \citenamefont
			{Tretiakov}, \citenamefont {Sharma}, \citenamefont {Fukami}, \citenamefont
			{Ohno},\ and\ \citenamefont {Singh}}]{chowdhury_unconventional_2021}%
		\BibitemOpen
		\bibfield  {author} {\bibinfo {author} {\bibfnamefont {R.~R.}\ \bibnamefont
				{Chowdhury}}, \bibinfo {author} {\bibfnamefont {S.}~\bibnamefont
				{DuttaGupta}}, \bibinfo {author} {\bibfnamefont {C.}~\bibnamefont {Patra}},
			\bibinfo {author} {\bibfnamefont {O.~A.}\ \bibnamefont {Tretiakov}}, \bibinfo
			{author} {\bibfnamefont {S.}~\bibnamefont {Sharma}}, \bibinfo {author}
			{\bibfnamefont {S.}~\bibnamefont {Fukami}}, \bibinfo {author} {\bibfnamefont
				{H.}~\bibnamefont {Ohno}},\ and\ \bibinfo {author} {\bibfnamefont {R.~P.}\
				\bibnamefont {Singh}},\ }\bibfield  {title} {\bibinfo {title} {Unconventional
				{Hall} effect and its variation with {Co}-doping in van der {Waals}
				{Fe$_3$GeTe$_2$}},\ }\href
		{https://www.nature.com/articles/s41598-021-93402-6} {\bibfield  {journal}
			{\bibinfo  {journal} {Scientific Reports}\ }\textbf {\bibinfo {volume}
				{11}},\ \bibinfo {pages} {14121} (\bibinfo {year} {2021})}\BibitemShut
		{NoStop}%
		\bibitem [{\citenamefont {Roy~Chowdhury}\ \emph {et~al.}(2022)\citenamefont
			{Roy~Chowdhury}, \citenamefont {Patra}, \citenamefont {DuttaGupta},
			\citenamefont {Satheesh}, \citenamefont {Dan}, \citenamefont {Fukami},\ and\
			\citenamefont {Singh}}]{roy_chowdhury_modification_2022}%
		\BibitemOpen
		\bibfield  {author} {\bibinfo {author} {\bibfnamefont {R.}~\bibnamefont
				{Roy~Chowdhury}}, \bibinfo {author} {\bibfnamefont {C.}~\bibnamefont
				{Patra}}, \bibinfo {author} {\bibfnamefont {S.}~\bibnamefont {DuttaGupta}},
			\bibinfo {author} {\bibfnamefont {S.}~\bibnamefont {Satheesh}}, \bibinfo
			{author} {\bibfnamefont {S.}~\bibnamefont {Dan}}, \bibinfo {author}
			{\bibfnamefont {S.}~\bibnamefont {Fukami}},\ and\ \bibinfo {author}
			{\bibfnamefont {R.~P.}\ \bibnamefont {Singh}},\ }\bibfield  {title} {\bibinfo
			{title} {Modification of unconventional {Hall} effect with doping at the
				nonmagnetic site in a two-dimensional van der {Waals} ferromagnet},\ }\href
		{https://link.aps.org/doi/10.1103/PhysRevMaterials.6.014002} {\bibfield
			{journal} {\bibinfo  {journal} {Phys. Rev. Materials}\ }\textbf {\bibinfo
				{volume} {6}},\ \bibinfo {pages} {014002} (\bibinfo {year}
			{2022})}\BibitemShut {NoStop}%
		\bibitem [{\citenamefont {Zhuang}\ \emph {et~al.}(2016)\citenamefont {Zhuang},
			\citenamefont {Kent},\ and\ \citenamefont {Hennig}}]{zhuang_strong_2016}%
		\BibitemOpen
		\bibfield  {author} {\bibinfo {author} {\bibfnamefont {H.~L.}\ \bibnamefont
				{Zhuang}}, \bibinfo {author} {\bibfnamefont {P.~R.~C.}\ \bibnamefont
				{Kent}},\ and\ \bibinfo {author} {\bibfnamefont {R.~G.}\ \bibnamefont
				{Hennig}},\ }\bibfield  {title} {\bibinfo {title} {Strong anisotropy and
				magnetostriction in the two-dimensional {Stoner} ferromagnet
				{Fe$_3$GeTe$_2$}},\ }\href
		{https://link.aps.org/doi/10.1103/PhysRevB.93.134407} {\bibfield  {journal}
			{\bibinfo  {journal} {Phys. Rev. B}\ }\textbf {\bibinfo {volume} {93}},\
			\bibinfo {pages} {134407} (\bibinfo {year} {2016})}\BibitemShut {NoStop}%
		\bibitem [{\citenamefont {Yuan}\ \emph {et~al.}(2017)\citenamefont {Yuan},
			\citenamefont {Jin}, \citenamefont {Liu}, \citenamefont {Shen}, \citenamefont
			{Lin}, \citenamefont {Li},\ and\ \citenamefont {Chen}}]{yuan_tuning_2017}%
		\BibitemOpen
		\bibfield  {author} {\bibinfo {author} {\bibfnamefont {D.}~\bibnamefont
				{Yuan}}, \bibinfo {author} {\bibfnamefont {S.}~\bibnamefont {Jin}}, \bibinfo
			{author} {\bibfnamefont {N.}~\bibnamefont {Liu}}, \bibinfo {author}
			{\bibfnamefont {S.}~\bibnamefont {Shen}}, \bibinfo {author} {\bibfnamefont
				{Z.}~\bibnamefont {Lin}}, \bibinfo {author} {\bibfnamefont {K.}~\bibnamefont
				{Li}},\ and\ \bibinfo {author} {\bibfnamefont {X.}~\bibnamefont {Chen}},\
		}\bibfield  {title} {\bibinfo
			{title} {Tuning magnetic properties in
				quasi-two-dimensional ferromagnetic {Fe$_{3-y}$Ge$_{1-x}$As$_x$Te$_2$ ($0~\leq~x~\leq~0.85$)}},\ }\href {https://iopscience.iop.org/article/10.1088/2053-1591/aa63b5} {\bibfield
			{journal} {\bibinfo  {journal} {Materials Research Express}\ }\textbf
			{\bibinfo {volume} {4}},\ \bibinfo {pages} {036103} (\bibinfo {year}
			{2017})}\BibitemShut {NoStop}%
		\bibitem [{\citenamefont {Zhang}\ \emph {et~al.}(2018)\citenamefont {Zhang},
			\citenamefont {Lu}, \citenamefont {Zhu}, \citenamefont {Tan}, \citenamefont
			{Feng}, \citenamefont {Liu}, \citenamefont {Zhang}, \citenamefont {Chen},
			\citenamefont {Liu}, \citenamefont {Luo}, \citenamefont {Xie}, \citenamefont
			{Luo}, \citenamefont {Zhang},\ and\ \citenamefont
			{Lai}}]{zhang_emergence_2018}%
		\BibitemOpen
		\bibfield  {author} {\bibinfo {author} {\bibfnamefont {Y.}~\bibnamefont
				{Zhang}}, \bibinfo {author} {\bibfnamefont {H.}~\bibnamefont {Lu}}, \bibinfo
			{author} {\bibfnamefont {X.}~\bibnamefont {Zhu}}, \bibinfo {author}
			{\bibfnamefont {S.}~\bibnamefont {Tan}}, \bibinfo {author} {\bibfnamefont
				{W.}~\bibnamefont {Feng}}, \bibinfo {author} {\bibfnamefont {Q.}~\bibnamefont
				{Liu}}, \bibinfo {author} {\bibfnamefont {W.}~\bibnamefont {Zhang}}, \bibinfo
			{author} {\bibfnamefont {Q.}~\bibnamefont {Chen}}, \bibinfo {author}
			{\bibfnamefont {Y.}~\bibnamefont {Liu}}, \bibinfo {author} {\bibfnamefont
				{X.}~\bibnamefont {Luo}}, \bibinfo {author} {\bibfnamefont {D.}~\bibnamefont
				{Xie}}, \bibinfo {author} {\bibfnamefont {L.}~\bibnamefont {Luo}}, \bibinfo
			{author} {\bibfnamefont {Z.}~\bibnamefont {Zhang}},\ and\ \bibinfo {author}
			{\bibfnamefont {X.}~\bibnamefont {Lai}},\ }\bibfield  {title} {\bibinfo
			{title} {Emergence of {Kondo} lattice behavior in a van der {Waals} itinerant
				ferromagnet, {Fe$_3$GeTe$_2$}},\ }\href
		{https://www.science.org/doi/10.1126/sciadv.aao6791} {\bibfield  {journal}
			{\bibinfo  {journal} {Science Advances}\ }\textbf {\bibinfo {volume} {4}},\
			\bibinfo {pages} {eaao6791} (\bibinfo {year} {2018})}\BibitemShut {NoStop}%
		\bibitem [{\citenamefont {Xu}\ \emph {et~al.}(2020)\citenamefont {Xu},
			\citenamefont {Li}, \citenamefont {Duan}, \citenamefont {Zhang},
			\citenamefont {Chen}, \citenamefont {Kang}, \citenamefont {Liang},
			\citenamefont {Chen}, \citenamefont {Xia}, \citenamefont {Xu}, \citenamefont
			{Malinowski}, \citenamefont {Xu}, \citenamefont {Chu}, \citenamefont {Li},
			\citenamefont {Guo}, \citenamefont {Liu}, \citenamefont {Yang},\ and\
			\citenamefont {Chen}}]{xu_signature_2020}%
		\BibitemOpen
		\bibfield  {author} {\bibinfo {author} {\bibfnamefont {X.}~\bibnamefont
				{Xu}}, \bibinfo {author} {\bibfnamefont {Y.~W.}\ \bibnamefont {Li}}, \bibinfo
			{author} {\bibfnamefont {S.~R.}\ \bibnamefont {Duan}}, \bibinfo {author}
			{\bibfnamefont {S.~L.}\ \bibnamefont {Zhang}}, \bibinfo {author}
			{\bibfnamefont {Y.~J.}\ \bibnamefont {Chen}}, \bibinfo {author}
			{\bibfnamefont {L.}~\bibnamefont {Kang}}, \bibinfo {author} {\bibfnamefont
				{A.~J.}\ \bibnamefont {Liang}}, \bibinfo {author} {\bibfnamefont
				{C.}~\bibnamefont {Chen}}, \bibinfo {author} {\bibfnamefont {W.}~\bibnamefont
				{Xia}}, \bibinfo {author} {\bibfnamefont {Y.}~\bibnamefont {Xu}}, \bibinfo
			{author} {\bibfnamefont {P.}~\bibnamefont {Malinowski}}, \bibinfo {author}
			{\bibfnamefont {X.~D.}\ \bibnamefont {Xu}}, \bibinfo {author} {\bibfnamefont
				{J.-H.}\ \bibnamefont {Chu}}, \bibinfo {author} {\bibfnamefont
				{G.}~\bibnamefont {Li}}, \bibinfo {author} {\bibfnamefont {Y.~F.}\
				\bibnamefont {Guo}}, \bibinfo {author} {\bibfnamefont {Z.~K.}\ \bibnamefont
				{Liu}}, \bibinfo {author} {\bibfnamefont {L.~X.}\ \bibnamefont {Yang}},\ and\
			\bibinfo {author} {\bibfnamefont {Y.~L.}\ \bibnamefont {Chen}},\ }\bibfield
		{title} {\bibinfo {title} {Signature for non-{Stoner} ferromagnetism in the
				van der {Waals} ferromagnet {Fe$_3$GeTe$_2$}},\ }\href
		{https://link.aps.org/doi/10.1103/PhysRevB.101.201104} {\bibfield  {journal}
			{\bibinfo  {journal} {Phys. Rev. B}\ }\textbf {\bibinfo {volume} {101}},\
			\bibinfo {pages} {201104} (\bibinfo {year} {2020})}\BibitemShut {NoStop}%
		\bibitem [{\citenamefont {Santiago}\ \emph {et~al.}(2017)\citenamefont
			{Santiago}, \citenamefont {Huang},\ and\ \citenamefont
			{Morosan}}]{santiago_itinerant_2017}%
		\BibitemOpen
		\bibfield  {author} {\bibinfo {author} {\bibfnamefont {J.~M.}\ \bibnamefont
				{Santiago}}, \bibinfo {author} {\bibfnamefont {C.-L.}\ \bibnamefont
				{Huang}},\ and\ \bibinfo {author} {\bibfnamefont {E.}~\bibnamefont
				{Morosan}},\ }\bibfield  {title} {\bibinfo {title} {Itinerant magnetic
				metals},\ }\href {https://dx.doi.org/10.1088/1361-648X/aa7889} {\bibfield
			{journal} {\bibinfo  {journal} {J. Phys.: Condens. Matter}\ }\textbf
			{\bibinfo {volume} {29}},\ \bibinfo {pages} {373002} (\bibinfo {year}
			{2017})}\BibitemShut {NoStop}%
		\bibitem [{\citenamefont {Korenman}\ \emph {et~al.}(1977)\citenamefont
			{Korenman}, \citenamefont {Murray},\ and\ \citenamefont
			{Prange}}]{korenman_local-band_1977}%
		\BibitemOpen
		\bibfield  {author} {\bibinfo {author} {\bibfnamefont {V.}~\bibnamefont
				{Korenman}}, \bibinfo {author} {\bibfnamefont {J.~L.}\ \bibnamefont
				{Murray}},\ and\ \bibinfo {author} {\bibfnamefont {R.~E.}\ \bibnamefont
				{Prange}},\ }\bibfield  {title} {\bibinfo {title} {Local-band theory of
				itinerant ferromagnetism. {I}. {Fermi}-liquid theory},\ }\href
		{https://link.aps.org/doi/10.1103/PhysRevB.16.4032} {\bibfield  {journal}
			{\bibinfo  {journal} {Phys. Rev. B}\ }\textbf {\bibinfo {volume} {16}},\
			\bibinfo {pages} {4032} (\bibinfo {year} {1977})}\BibitemShut {NoStop}%
		\bibitem [{\citenamefont {Maiti}\ \emph {et~al.}(2002)\citenamefont {Maiti},
			\citenamefont {Malagoli}, \citenamefont {Dallmeyer},\ and\ \citenamefont
			{Carbone}}]{maiti_finite_2002}%
		\BibitemOpen
		\bibfield  {author} {\bibinfo {author} {\bibfnamefont {K.}~\bibnamefont
				{Maiti}}, \bibinfo {author} {\bibfnamefont {M.~C.}\ \bibnamefont {Malagoli}},
			\bibinfo {author} {\bibfnamefont {A.}~\bibnamefont {Dallmeyer}},\ and\
			\bibinfo {author} {\bibfnamefont {C.}~\bibnamefont {Carbone}},\ }\bibfield
		{title} {\bibinfo {title} {Finite {Temperature} {Magnetism} in {Gd}:
				{Evidence} against a {Stoner} {Behavior}},\ }\href
		{https://link.aps.org/doi/10.1103/PhysRevLett.88.167205} {\bibfield
			{journal} {\bibinfo  {journal} {Phys. Rev. Lett.}\ }\textbf {\bibinfo
				{volume} {88}},\ \bibinfo {pages} {167205} (\bibinfo {year}
			{2002})}\BibitemShut {NoStop}%
		\bibitem [{\citenamefont {Johnston}(2010)}]{johnston_puzzle_2010}%
		\BibitemOpen
		\bibfield  {author} {\bibinfo {author} {\bibfnamefont {D.~C.}\ \bibnamefont
				{Johnston}},\ }\bibfield  {title} {\bibinfo {title} {The puzzle of high
				temperature superconductivity in layered iron pnictides and chalcogenides},\
		}\href {https://doi.org/10.1080/00018732.2010.513480} {\bibfield  {journal}
			{\bibinfo  {journal} {Advances in Physics}\ }\textbf {\bibinfo {volume}
				{59}},\ \bibinfo {pages} {803} (\bibinfo {year} {2010})}\BibitemShut
		{NoStop}%
		\bibitem [{\citenamefont {Hansmann}\ \emph {et~al.}(2010)\citenamefont
			{Hansmann}, \citenamefont {Arita}, \citenamefont {Toschi}, \citenamefont
			{Sakai}, \citenamefont {Sangiovanni},\ and\ \citenamefont
			{Held}}]{Hansmann_LaFeAsODMFT_PRL2010}%
		\BibitemOpen
		\bibfield  {author} {\bibinfo {author} {\bibfnamefont {P.}~\bibnamefont
				{Hansmann}}, \bibinfo {author} {\bibfnamefont {R.}~\bibnamefont {Arita}},
			\bibinfo {author} {\bibfnamefont {A.}~\bibnamefont {Toschi}}, \bibinfo
			{author} {\bibfnamefont {S.}~\bibnamefont {Sakai}}, \bibinfo {author}
			{\bibfnamefont {G.}~\bibnamefont {Sangiovanni}},\ and\ \bibinfo {author}
			{\bibfnamefont {K.}~\bibnamefont {Held}},\ }\bibfield  {title} {\bibinfo
			{title} {Dichotomy between large local and small ordered magnetic moments in
				iron-based superconductors},\ }\href
		{https://link.aps.org/doi/10.1103/PhysRevLett.104.197002} {\bibfield
			{journal} {\bibinfo  {journal} {Phys. Rev. Lett.}\ }\textbf {\bibinfo
				{volume} {104}},\ \bibinfo {pages} {197002} (\bibinfo {year}
			{2010})}\BibitemShut {NoStop}%
		\bibitem [{\citenamefont {Dai}\ \emph {et~al.}(2012)\citenamefont {Dai},
			\citenamefont {Hu},\ and\ \citenamefont {Dagotto}}]{dai_magnetism_2012}%
		\BibitemOpen
		\bibfield  {author} {\bibinfo {author} {\bibfnamefont {P.}~\bibnamefont
				{Dai}}, \bibinfo {author} {\bibfnamefont {J.}~\bibnamefont {Hu}},\ and\
			\bibinfo {author} {\bibfnamefont {E.}~\bibnamefont {Dagotto}},\ }\bibfield
		{title} {\bibinfo {title} {Magnetism and its microscopic origin in iron-based
				high-temperature superconductors},\ }\href
		{https://www.nature.com/articles/nphys2438} {\bibfield  {journal} {\bibinfo
				{journal} {Nature Physics}\ }\textbf {\bibinfo {volume} {8}},\ \bibinfo
			{pages} {709} (\bibinfo {year} {2012})}\BibitemShut {NoStop}%
		\bibitem [{\citenamefont {Haule}\ and\ \citenamefont
			{Kotliar}(2009)}]{haule_coherenceincoherence_2009}%
		\BibitemOpen
		\bibfield  {author} {\bibinfo {author} {\bibfnamefont {K.}~\bibnamefont
				{Haule}}\ and\ \bibinfo {author} {\bibfnamefont {G.}~\bibnamefont
				{Kotliar}},\ }\bibfield  {title} {\bibinfo {title} {Coherence–incoherence
				crossover in the normal state of iron oxypnictides and importance of {Hund}'s
				rule coupling},\ }\href {https://dx.doi.org/10.1088/1367-2630/11/2/025021}
		{\bibfield  {journal} {\bibinfo  {journal} {New Journal of Physics}\ }\textbf
			{\bibinfo {volume} {11}},\ \bibinfo {pages} {025021} (\bibinfo {year}
			{2009})}\BibitemShut {NoStop}%
		\bibitem [{\citenamefont {Bai}\ \emph {et~al.}(2022)\citenamefont {Bai},
			\citenamefont {Lechermann}, \citenamefont {Liu}, \citenamefont {Cheng},
			\citenamefont {Kolesnikov}, \citenamefont {Ye}, \citenamefont {Williams},
			\citenamefont {Chi}, \citenamefont {Hong}, \citenamefont {Granroth},
			\citenamefont {May},\ and\ \citenamefont {Calder}}]{AFM}%
		\BibitemOpen
		\bibfield  {author} {\bibinfo {author} {\bibfnamefont {X.}~\bibnamefont
				{Bai}}, \bibinfo {author} {\bibfnamefont {F.}~\bibnamefont {Lechermann}},
			\bibinfo {author} {\bibfnamefont {Y.}~\bibnamefont {Liu}}, \bibinfo {author}
			{\bibfnamefont {Y.}~\bibnamefont {Cheng}}, \bibinfo {author} {\bibfnamefont
				{A.~I.}\ \bibnamefont {Kolesnikov}}, \bibinfo {author} {\bibfnamefont
				{F.}~\bibnamefont {Ye}}, \bibinfo {author} {\bibfnamefont {T.~J.}\
				\bibnamefont {Williams}}, \bibinfo {author} {\bibfnamefont {S.}~\bibnamefont
				{Chi}}, \bibinfo {author} {\bibfnamefont {T.}~\bibnamefont {Hong}}, \bibinfo
			{author} {\bibfnamefont {G.~E.}\ \bibnamefont {Granroth}}, \bibinfo {author}
			{\bibfnamefont {A.~F.}\ \bibnamefont {May}},\ and\ \bibinfo {author}
			{\bibfnamefont {S.}~\bibnamefont {Calder}},\ }\bibfield  {title} {\bibinfo
			{title} {Antiferromagnetic fluctuations and orbital-selective mott transition
				in the van der waals ferromagnet {Fe$_3$GeTe$_2$}},\ }\href@noop {}
		{\bibfield  {journal} {\bibinfo  {journal} {Phys. Rev. B}\ }\textbf {\bibinfo
				{volume} {106}},\ \bibinfo {pages} {L180409} (\bibinfo {year}
			{2022})}\BibitemShut {NoStop}%
		\bibitem [{\citenamefont {Zhao}\ \emph {et~al.}(2021)\citenamefont {Zhao},
			\citenamefont {Chen}, \citenamefont {Xi}, \citenamefont {Zhao}, \citenamefont
			{Xu}, \citenamefont {Zhang}, \citenamefont {Cheng}, \citenamefont {Feng},
			\citenamefont {Zhuang}, \citenamefont {Pan}, \citenamefont {Xu},
			\citenamefont {Hao}, \citenamefont {Li}, \citenamefont {Zhou}, \citenamefont
			{Dou},\ and\ \citenamefont {Du}}]{zhao_kondo_2021}%
		\BibitemOpen
		\bibfield  {author} {\bibinfo {author} {\bibfnamefont {M.}~\bibnamefont
				{Zhao}}, \bibinfo {author} {\bibfnamefont {B.-B.}\ \bibnamefont {Chen}},
			\bibinfo {author} {\bibfnamefont {Y.}~\bibnamefont {Xi}}, \bibinfo {author}
			{\bibfnamefont {Y.}~\bibnamefont {Zhao}}, \bibinfo {author} {\bibfnamefont
				{H.}~\bibnamefont {Xu}}, \bibinfo {author} {\bibfnamefont {H.}~\bibnamefont
				{Zhang}}, \bibinfo {author} {\bibfnamefont {N.}~\bibnamefont {Cheng}},
			\bibinfo {author} {\bibfnamefont {H.}~\bibnamefont {Feng}}, \bibinfo {author}
			{\bibfnamefont {J.}~\bibnamefont {Zhuang}}, \bibinfo {author} {\bibfnamefont
				{F.}~\bibnamefont {Pan}}, \bibinfo {author} {\bibfnamefont {X.}~\bibnamefont
				{Xu}}, \bibinfo {author} {\bibfnamefont {W.}~\bibnamefont {Hao}}, \bibinfo
			{author} {\bibfnamefont {W.}~\bibnamefont {Li}}, \bibinfo {author}
			{\bibfnamefont {S.}~\bibnamefont {Zhou}}, \bibinfo {author} {\bibfnamefont
				{S.~X.}\ \bibnamefont {Dou}},\ and\ \bibinfo {author} {\bibfnamefont
				{Y.}~\bibnamefont {Du}},\ }\bibfield  {title} {\bibinfo {title} {Kondo
				{Holes} in the {Two}-{Dimensional} {Itinerant} {Ising} {Ferromagnet}
				{Fe$_3$GeTe$_2$}},\ }\href {https://doi.org/10.1021/acs.nanolett.1c01661}
		{\bibfield  {journal} {\bibinfo  {journal} {Nano Letters}\ }\textbf {\bibinfo
				{volume} {21}},\ \bibinfo {pages} {6117} (\bibinfo {year}
			{2021})}\BibitemShut {NoStop}%
		\bibitem [{\citenamefont {Rana}\ \emph {et~al.}(2022)\citenamefont {Rana},
			\citenamefont {R}, \citenamefont {G}, \citenamefont {Patra}, \citenamefont
			{Howlader}, \citenamefont {Chowdhury}, \citenamefont {Kabir}, \citenamefont
			{Singh},\ and\ \citenamefont {Sheet}}]{rana_spin-polarized_2022}%
		\BibitemOpen
		\bibfield  {author} {\bibinfo {author} {\bibfnamefont {D.}~\bibnamefont
				{Rana}}, \bibinfo {author} {\bibfnamefont {A.}~\bibnamefont {R}}, \bibinfo
			{author} {\bibfnamefont {B.}~\bibnamefont {G}}, \bibinfo {author}
			{\bibfnamefont {C.}~\bibnamefont {Patra}}, \bibinfo {author} {\bibfnamefont
				{S.}~\bibnamefont {Howlader}}, \bibinfo {author} {\bibfnamefont {R.~R.}\
				\bibnamefont {Chowdhury}}, \bibinfo {author} {\bibfnamefont {M.}~\bibnamefont
				{Kabir}}, \bibinfo {author} {\bibfnamefont {R.~P.}\ \bibnamefont {Singh}},\
			and\ \bibinfo {author} {\bibfnamefont {G.}~\bibnamefont {Sheet}},\ }\bibfield
		{title} {\bibinfo {title} {Spin-polarized supercurrent through the van der
				{Waals} {Kondo}-lattice ferromagnet {Fe$_3$GeTe$_2$}},\ }\href
		{https://link.aps.org/doi/10.1103/PhysRevB.106.085120} {\bibfield  {journal}
			{\bibinfo  {journal} {Phys. Rev. B}\ }\textbf {\bibinfo {volume} {106}},\
			\bibinfo {pages} {085120} (\bibinfo {year} {2022})}\BibitemShut {NoStop}%
		\bibitem [{\citenamefont {Zhu}\ \emph {et~al.}(2016)\citenamefont {Zhu},
			\citenamefont {Janoschek}, \citenamefont {Chaves}, \citenamefont {Cezar},
			\citenamefont {Durakiewicz}, \citenamefont {Ronning}, \citenamefont {Sassa},
			\citenamefont {Mansson}, \citenamefont {Scott}, \citenamefont {Wakeham},
			\citenamefont {Bauer},\ and\ \citenamefont {Thompson}}]{zhu_electronic_2016}%
		\BibitemOpen
		\bibfield  {author} {\bibinfo {author} {\bibfnamefont {J.-X.}\ \bibnamefont
				{Zhu}}, \bibinfo {author} {\bibfnamefont {M.}~\bibnamefont {Janoschek}},
			\bibinfo {author} {\bibfnamefont {D.~S.}\ \bibnamefont {Chaves}}, \bibinfo
			{author} {\bibfnamefont {J.~C.}\ \bibnamefont {Cezar}}, \bibinfo {author}
			{\bibfnamefont {T.}~\bibnamefont {Durakiewicz}}, \bibinfo {author}
			{\bibfnamefont {F.}~\bibnamefont {Ronning}}, \bibinfo {author} {\bibfnamefont
				{Y.}~\bibnamefont {Sassa}}, \bibinfo {author} {\bibfnamefont
				{M.}~\bibnamefont {Mansson}}, \bibinfo {author} {\bibfnamefont {B.~L.}\
				\bibnamefont {Scott}}, \bibinfo {author} {\bibfnamefont {N.}~\bibnamefont
				{Wakeham}}, \bibinfo {author} {\bibfnamefont {E.~D.}\ \bibnamefont {Bauer}},\
			and\ \bibinfo {author} {\bibfnamefont {J.~D.}\ \bibnamefont {Thompson}},\
		}\bibfield  {title} {\bibinfo {title} {Electronic correlation and magnetism
				in the ferromagnetic metal {Fe$_3$GeTe$_2$}},\ }\href
		{https://link.aps.org/doi/10.1103/PhysRevB.93.144404} {\bibfield  {journal}
			{\bibinfo  {journal} {Phys. Rev. B}\ }\textbf {\bibinfo {volume} {93}},\
			\bibinfo {pages} {144404} (\bibinfo {year} {2016})}\BibitemShut {NoStop}%
		\bibitem [{\citenamefont {Kim}\ \emph {et~al.}(2018)\citenamefont {Kim},
			\citenamefont {Seo}, \citenamefont {Lee}, \citenamefont {Ko}, \citenamefont
			{Kim}, \citenamefont {Jang}, \citenamefont {Ok}, \citenamefont {Lee},
			\citenamefont {Jo}, \citenamefont {Kang}, \citenamefont {Shim}, \citenamefont
			{Kim}, \citenamefont {Yeom}, \citenamefont {Il~Min}, \citenamefont {Yang},\
			and\ \citenamefont {Kim}}]{kim_large_2018}%
		\BibitemOpen
		\bibfield  {author} {\bibinfo {author} {\bibfnamefont {K.}~\bibnamefont
				{Kim}}, \bibinfo {author} {\bibfnamefont {J.}~\bibnamefont {Seo}}, \bibinfo
			{author} {\bibfnamefont {E.}~\bibnamefont {Lee}}, \bibinfo {author}
			{\bibfnamefont {K.-T.}\ \bibnamefont {Ko}}, \bibinfo {author} {\bibfnamefont
				{B.~S.}\ \bibnamefont {Kim}}, \bibinfo {author} {\bibfnamefont {B.~G.}\
				\bibnamefont {Jang}}, \bibinfo {author} {\bibfnamefont {J.~M.}\ \bibnamefont
				{Ok}}, \bibinfo {author} {\bibfnamefont {J.}~\bibnamefont {Lee}}, \bibinfo
			{author} {\bibfnamefont {Y.~J.}\ \bibnamefont {Jo}}, \bibinfo {author}
			{\bibfnamefont {W.}~\bibnamefont {Kang}}, \bibinfo {author} {\bibfnamefont
				{J.~H.}\ \bibnamefont {Shim}}, \bibinfo {author} {\bibfnamefont
				{C.}~\bibnamefont {Kim}}, \bibinfo {author} {\bibfnamefont {H.~W.}\
				\bibnamefont {Yeom}}, \bibinfo {author} {\bibfnamefont {B.}~\bibnamefont
				{Il~Min}}, \bibinfo {author} {\bibfnamefont {B.-J.}\ \bibnamefont {Yang}},\
			and\ \bibinfo {author} {\bibfnamefont {J.~S.}\ \bibnamefont {Kim}},\
		}\bibfield  {title} {\bibinfo {title} {Large anomalous {Hall} current induced
				by topological nodal lines in a ferromagnetic van der {Waals} semimetal},\
		}\href {https://www.nature.com/articles/s41563-018-0132-3} {\bibfield
			{journal} {\bibinfo  {journal} {Nature Materials}\ }\textbf {\bibinfo
				{volume} {17}},\ \bibinfo {pages} {794} (\bibinfo {year} {2018})}\BibitemShut
		{NoStop}%
		\bibitem [{SM()}]{SM}%
		\BibitemOpen
		\href@noop {} {\bibinfo  {journal} {See Supplemental Material at [URL will be
				inserted by publisher] for experimental details and analysis and for further
				DFT and DFT+DMFT results.}\ }\BibitemShut {NoStop}%
		\bibitem [{\citenamefont {Rossi}\ \emph {et~al.}(1997)\citenamefont {Rossi},
			\citenamefont {Panaccione}, \citenamefont {Sirotti}, \citenamefont {Lizzit},
			\citenamefont {Baraldi},\ and\ \citenamefont {Paolucci}}]{PhysRevB.55.11488}%
		\BibitemOpen
		\bibfield  {journal} {  }\bibfield  {author} {\bibinfo {author} {\bibfnamefont
				{G.}~\bibnamefont {Rossi}}, \bibinfo {author} {\bibfnamefont
				{G.}~\bibnamefont {Panaccione}}, \bibinfo {author} {\bibfnamefont
				{F.}~\bibnamefont {Sirotti}}, \bibinfo {author} {\bibfnamefont
				{S.}~\bibnamefont {Lizzit}}, \bibinfo {author} {\bibfnamefont
				{A.}~\bibnamefont {Baraldi}},\ and\ \bibinfo {author} {\bibfnamefont
				{G.}~\bibnamefont {Paolucci}},\ }\bibfield  {title} {\bibinfo {title}
			{Magnetic dichroism in the angular distribution of {F}e 2$p$ and 3$p$
				photoelectrons: Empirical support to zeeman-like analysis},\ }\href
		{https://doi.org/10.1103/PhysRevB.55.11488} {\bibfield  {journal} {\bibinfo
				{journal} {Phys. Rev. B}\ }\textbf {\bibinfo {volume} {55}},\ \bibinfo
			{pages} {11488} (\bibinfo {year} {1997})}\BibitemShut {NoStop}%
		\bibitem [{\citenamefont {Chen}\ \emph {et~al.}(2013)\citenamefont {Chen},
			\citenamefont {Yang}, \citenamefont {Wang}, \citenamefont {Imai},
			\citenamefont {Ohta}, \citenamefont {Michioka}, \citenamefont {Yoshimura},\
			and\ \citenamefont {Fang}}]{chen_magnetic_2013}%
		\BibitemOpen
		\bibfield  {author} {\bibinfo {author} {\bibfnamefont {B.}~\bibnamefont
				{Chen}}, \bibinfo {author} {\bibfnamefont {J.}~\bibnamefont {Yang}}, \bibinfo
			{author} {\bibfnamefont {H.}~\bibnamefont {Wang}}, \bibinfo {author}
			{\bibfnamefont {M.}~\bibnamefont {Imai}}, \bibinfo {author} {\bibfnamefont
				{H.}~\bibnamefont {Ohta}}, \bibinfo {author} {\bibfnamefont {C.}~\bibnamefont
				{Michioka}}, \bibinfo {author} {\bibfnamefont {K.}~\bibnamefont
				{Yoshimura}},\ and\ \bibinfo {author} {\bibfnamefont {M.}~\bibnamefont
				{Fang}},\ }\bibfield  {title} {\bibinfo {title} {Magnetic {Properties} of
				{Layered} {Itinerant} {Electron} {Ferromagnet} {Fe$_3$GeTe$_2$}},\ }\href
		{https://journals.jps.jp/doi/abs/10.7566/JPSJ.82.124711} {\bibfield
			{journal} {\bibinfo  {journal} {J. Phys. Soc. Jpn.}\ }\textbf {\bibinfo
				{volume} {82}},\ \bibinfo {pages} {124711} (\bibinfo {year}
			{2013})}\BibitemShut {NoStop}%
		\bibitem [{\citenamefont {Blaha}\ \emph {et~al.}(2020)\citenamefont {Blaha},
			\citenamefont {Schwarz}, \citenamefont {Tran}, \citenamefont {Laskowski},
			\citenamefont {Madsen},\ and\ \citenamefont {Marks}}]{Wien2k2020}%
		\BibitemOpen
		\bibfield  {author} {\bibinfo {author} {\bibfnamefont {P.}~\bibnamefont
				{Blaha}}, \bibinfo {author} {\bibfnamefont {K.}~\bibnamefont {Schwarz}},
			\bibinfo {author} {\bibfnamefont {F.}~\bibnamefont {Tran}}, \bibinfo {author}
			{\bibfnamefont {R.}~\bibnamefont {Laskowski}}, \bibinfo {author}
			{\bibfnamefont {G.~K.~H.}\ \bibnamefont {Madsen}},\ and\ \bibinfo {author}
			{\bibfnamefont {L.~D.}\ \bibnamefont {Marks}},\ }\bibfield  {title} {\bibinfo
			{title} {Wien2k: An {APW}+lo program for calculating the properties of
				solids},\ }\href {https://doi.org/10.1063/1.5143061} {\bibfield  {journal}
			{\bibinfo  {journal} {J. Chem. Phys.}\ }\textbf {\bibinfo {volume} {152}},\
			\bibinfo {pages} {074101} (\bibinfo {year} {2020})}\BibitemShut {NoStop}%
		\bibitem [{\citenamefont {Perdew}\ \emph {et~al.}(1996)\citenamefont {Perdew},
			\citenamefont {Burke},\ and\ \citenamefont
			{Ernzerhof}}]{perdew_generalized_1996}%
		\BibitemOpen
		\bibfield  {author} {\bibinfo {author} {\bibfnamefont {J.~P.}\ \bibnamefont
				{Perdew}}, \bibinfo {author} {\bibfnamefont {K.}~\bibnamefont {Burke}},\ and\
			\bibinfo {author} {\bibfnamefont {M.}~\bibnamefont {Ernzerhof}},\ }\bibfield
		{title} {\bibinfo {title} {Generalized {Gradient} {Approximation} {Made}
				{Simple}},\ }\href {https://link.aps.org/doi/10.1103/PhysRevLett.77.3865}
		{\bibfield  {journal} {\bibinfo  {journal} {Phys. Rev. Lett.}\ }\textbf
			{\bibinfo {volume} {77}},\ \bibinfo {pages} {3865} (\bibinfo {year}
			{1996})}\BibitemShut {NoStop}%
		\bibitem [{\citenamefont {Haule}\ \emph {et~al.}(2010)\citenamefont {Haule},
			\citenamefont {Yee},\ and\ \citenamefont {Kim}}]{haule_dynamical_2010}%
		\BibitemOpen
		\bibfield  {author} {\bibinfo {author} {\bibfnamefont {K.}~\bibnamefont
				{Haule}}, \bibinfo {author} {\bibfnamefont {C.-H.}\ \bibnamefont {Yee}},\
			and\ \bibinfo {author} {\bibfnamefont {K.}~\bibnamefont {Kim}},\ }\bibfield
		{title} {\bibinfo {title} {Dynamical mean-field theory within the
				full-potential methods: {Electronic} structure of {CeIrIn}$_5$ , {CeCoIn}$_5$
				, and {CeRhIn}$_5$},\ }\href
		{https://link.aps.org/doi/10.1103/PhysRevB.81.195107} {\bibfield  {journal}
			{\bibinfo  {journal} {Phys. Rev. B}\ }\textbf {\bibinfo {volume} {81}},\
			\bibinfo {pages} {195107} (\bibinfo {year} {2010})}\BibitemShut {NoStop}%
		\bibitem [{\citenamefont {Haule}(2007)}]{haule_quantum_2007}%
		\BibitemOpen
		\bibfield  {author} {\bibinfo {author} {\bibfnamefont {K.}~\bibnamefont
				{Haule}},\ }\bibfield  {title} {\bibinfo {title} {Quantum {Monte} {Carlo}
				impurity solver for cluster dynamical mean-field theory and electronic
				structure calculations with adjustable cluster base},\ }\href
		{https://link.aps.org/doi/10.1103/PhysRevB.75.155113} {\bibfield  {journal}
			{\bibinfo  {journal} {Phys. Rev. B}\ }\textbf {\bibinfo {volume} {75}},\
			\bibinfo {pages} {155113} (\bibinfo {year} {2007})}\BibitemShut {NoStop}%
		\bibitem [{\citenamefont {Haule}(2015)}]{haule_exact_2015}%
		\BibitemOpen
		\bibfield  {author} {\bibinfo {author} {\bibfnamefont {K.}~\bibnamefont
				{Haule}},\ }\bibfield  {title} {\bibinfo {title} {Exact {Double} {Counting}
				in {Combining} the {Dynamical} {Mean} {Field} {Theory} and the {Density}
				{Functional} {Theory}},\ }\href
		{https://link.aps.org/doi/10.1103/PhysRevLett.115.196403} {\bibfield
			{journal} {\bibinfo  {journal} {Phys. Rev. Lett.}\ }\textbf {\bibinfo
				{volume} {115}},\ \bibinfo {pages} {196403} (\bibinfo {year}
			{2015})}\BibitemShut {NoStop}%
		\bibitem [{\citenamefont {Moulder}\ and\ \citenamefont
			{Chastain}(1992)}]{Handbook}%
		\BibitemOpen
		\bibfield  {author} {\bibinfo {author} {\bibfnamefont {J.}~\bibnamefont
				{Moulder}}\ and\ \bibinfo {author} {\bibfnamefont {J.}~\bibnamefont
				{Chastain}},\ }\href {https://books.google.co.in/books?id=A_XGQgAACAAJ}
		{\emph {\bibinfo {title} {Handbook of X-ray Photoelectron Spectroscopy: A
					Reference Book of Standard Spectra for Identification and Interpretation of
					XPS Data}}}\ (\bibinfo  {publisher} {Physical Electronics Division,
			Perkin-Elmer Corporation},\ \bibinfo {year} {1992})\BibitemShut {NoStop}%
		\bibitem [{\citenamefont {Sarkar}\ \emph {et~al.}(2021)\citenamefont {Sarkar},
			\citenamefont {Sadhukhan}, \citenamefont {Singh}, \citenamefont
			{Gloskovskii}, \citenamefont {Deguchi}, \citenamefont {Fujita},\ and\
			\citenamefont {Barman}}]{PhysRevResearch.3.013151}%
		\BibitemOpen
		\bibfield  {author} {\bibinfo {author} {\bibfnamefont {S.}~\bibnamefont
				{Sarkar}}, \bibinfo {author} {\bibfnamefont {P.}~\bibnamefont {Sadhukhan}},
			\bibinfo {author} {\bibfnamefont {V.~K.}\ \bibnamefont {Singh}}, \bibinfo
			{author} {\bibfnamefont {A.}~\bibnamefont {Gloskovskii}}, \bibinfo {author}
			{\bibfnamefont {K.}~\bibnamefont {Deguchi}}, \bibinfo {author} {\bibfnamefont
				{N.}~\bibnamefont {Fujita}},\ and\ \bibinfo {author} {\bibfnamefont {S.~R.}\
				\bibnamefont {Barman}},\ }\bibfield  {title} {\bibinfo {title} {Bulk
				electronic structure of high-order quaternary approximants},\ }\href
		{https://doi.org/10.1103/PhysRevResearch.3.013151} {\bibfield  {journal}
			{\bibinfo  {journal} {Phys. Rev. Res.}\ }\textbf {\bibinfo {volume} {3}},\
			\bibinfo {pages} {013151} (\bibinfo {year} {2021})}\BibitemShut {NoStop}%
		\bibitem [{\citenamefont {Baumgarten}\ \emph {et~al.}(1990)\citenamefont
			{Baumgarten}, \citenamefont {Schneider}, \citenamefont {Petersen},
			\citenamefont {Sch\"afers},\ and\ \citenamefont
			{Kirschner}}]{PhysRevLett.65.492}%
		\BibitemOpen
		\bibfield  {author} {\bibinfo {author} {\bibfnamefont {L.}~\bibnamefont
				{Baumgarten}}, \bibinfo {author} {\bibfnamefont {C.~M.}\ \bibnamefont
				{Schneider}}, \bibinfo {author} {\bibfnamefont {H.}~\bibnamefont {Petersen}},
			\bibinfo {author} {\bibfnamefont {F.}~\bibnamefont {Sch\"afers}},\ and\
			\bibinfo {author} {\bibfnamefont {J.}~\bibnamefont {Kirschner}},\ }\bibfield
		{title} {\bibinfo {title} {Magnetic x-ray dichroism in core-level
				photoemission from ferromagnets},\ }\href
		{https://doi.org/10.1103/PhysRevLett.65.492} {\bibfield  {journal} {\bibinfo
				{journal} {Phys. Rev. Lett.}\ }\textbf {\bibinfo {volume} {65}},\ \bibinfo
			{pages} {492} (\bibinfo {year} {1990})}\BibitemShut {NoStop}%
		\bibitem [{\citenamefont {Menchero}(1998)}]{PhysRevB.57.993}%
		\BibitemOpen
		\bibfield  {author} {\bibinfo {author} {\bibfnamefont {J.~G.}\ \bibnamefont
				{Menchero}},\ }\bibfield  {title} {\bibinfo {title} {One-electron theory of
				core-level photoemission from ferromagnets},\ }\href
		{https://doi.org/10.1103/PhysRevB.57.993} {\bibfield  {journal} {\bibinfo
				{journal} {Phys. Rev. B}\ }\textbf {\bibinfo {volume} {57}},\ \bibinfo
			{pages} {993} (\bibinfo {year} {1998})}\BibitemShut {NoStop}%
		\bibitem [{\citenamefont {Yeh}\ and\ \citenamefont
			{Lindau}(1985)}]{yeh_atomic_1985}%
		\BibitemOpen
		\bibfield  {author} {\bibinfo {author} {\bibfnamefont {J.~J.}\ \bibnamefont
				{Yeh}}\ and\ \bibinfo {author} {\bibfnamefont {I.}~\bibnamefont {Lindau}},\
		}\bibfield  {title} {\bibinfo {title} {Atomic subshell photoionization cross
				sections and asymmetry parameters: 1 $\leq$ {Z} $\leq$ 103},\ }\href
		{https://www.sciencedirect.com/science/article/pii/0092640X85900166}
		{\bibfield  {journal} {\bibinfo  {journal} {Atomic Data and Nuclear Data
					Tables}\ }\textbf {\bibinfo {volume} {32}},\ \bibinfo {pages} {1} (\bibinfo
			{year} {1985})}\BibitemShut {NoStop}%
		\bibitem [{\citenamefont {Kim}\ and\ \citenamefont
			{Min}(2015)}]{kim_nature_2015}%
		\BibitemOpen
		\bibfield  {author} {\bibinfo {author} {\bibfnamefont {M.}~\bibnamefont
				{Kim}}\ and\ \bibinfo {author} {\bibfnamefont {B.~I.}\ \bibnamefont {Min}},\
		}\bibfield  {title} {\bibinfo {title} {Nature of itinerant ferromagnetism of
				{SrRuO}$_3$ : {A} {DFT}+{DMFT} study},\ }\href
		{https://link.aps.org/doi/10.1103/PhysRevB.91.205116} {\bibfield  {journal}
			{\bibinfo  {journal} {Phys. Rev. B}\ }\textbf {\bibinfo {volume} {91}},\
			\bibinfo {pages} {205116} (\bibinfo {year} {2015})}\BibitemShut {NoStop}%
		\bibitem [{\citenamefont {Singh}\ \emph {et~al.}(2007)\citenamefont {Singh},
			\citenamefont {Medicherla},\ and\ \citenamefont {Maiti}}]{10.1063/1.2789731}%
		\BibitemOpen
		\bibfield  {author} {\bibinfo {author} {\bibfnamefont {R.~S.}\ \bibnamefont
				{Singh}}, \bibinfo {author} {\bibfnamefont {V.~R.~R.}\ \bibnamefont
				{Medicherla}},\ and\ \bibinfo {author} {\bibfnamefont {K.}~\bibnamefont
				{Maiti}},\ }\bibfield  {title} {\bibinfo {title} {{Role of long range
					ferromagnetic order in the electronic structure of Sr$_{1 -
						x}$Ca$_x$RuO$_3$}},\ }\href {https://doi.org/10.1063/1.2789731} {\bibfield
			{journal} {\bibinfo  {journal} {Applied Physics Letters}\ }\textbf {\bibinfo
				{volume} {91}},\ \bibinfo {pages} {132503} (\bibinfo {year}
			{2007})}\BibitemShut {NoStop}%
		\bibitem [{\citenamefont {Ghosh}\ \emph {et~al.}(2023)\citenamefont {Ghosh},
			\citenamefont {Ershadrad}, \citenamefont {Borisov},\ and\ \citenamefont
			{Sanyal}}]{ghosh_unraveling_2023}%
		\BibitemOpen
		\bibfield  {author} {\bibinfo {author} {\bibfnamefont {S.}~\bibnamefont
				{Ghosh}}, \bibinfo {author} {\bibfnamefont {S.}~\bibnamefont {Ershadrad}},
			\bibinfo {author} {\bibfnamefont {V.}~\bibnamefont {Borisov}},\ and\ \bibinfo
			{author} {\bibfnamefont {B.}~\bibnamefont {Sanyal}},\ }\bibfield  {title}
		{\bibinfo {title} {Unraveling effects of electron correlation in
				two-dimensional {Fe$_n$GeTe$_2$} (n = 3, 4, 5) by dynamical mean field
				theory},\ }\href {https://www.nature.com/articles/s41524-023-01024-5}
		{\bibfield  {journal} {\bibinfo  {journal} {npj Computational Materials}\
			}\textbf {\bibinfo {volume} {9}},\ \bibinfo {pages} {1} (\bibinfo {year}
			{2023})}\BibitemShut {NoStop}%
		\bibitem [{\citenamefont {Kim}\ \emph {et~al.}(2022)\citenamefont {Kim},
			\citenamefont {Ryee},\ and\ \citenamefont {Han}}]{kim_fe3gete2_2022}%
		\BibitemOpen
		\bibfield  {author} {\bibinfo {author} {\bibfnamefont {T.~J.}\ \bibnamefont
				{Kim}}, \bibinfo {author} {\bibfnamefont {S.}~\bibnamefont {Ryee}},\ and\
			\bibinfo {author} {\bibfnamefont {M.~J.}\ \bibnamefont {Han}},\ }\bibfield
		{title} {\bibinfo {title} {{Fe$_3$GeTe$_2$}: a site-differentiated {Hund}
				metal},\ }\href {https://www.nature.com/articles/s41524-022-00937-x}
		{\bibfield  {journal} {\bibinfo  {journal} {npj Computational Materials}\
			}\textbf {\bibinfo {volume} {8}},\ \bibinfo {pages} {1} (\bibinfo {year}
			{2022})}\BibitemShut {NoStop}%
		\bibitem [{\citenamefont {Paul}\ and\ \citenamefont
			{Birol}(2019)}]{DFT+DMFTarpitapaul}%
		\BibitemOpen
		\bibfield  {author} {\bibinfo {author} {\bibfnamefont {A.}~\bibnamefont
				{Paul}}\ and\ \bibinfo {author} {\bibfnamefont {T.}~\bibnamefont {Birol}},\
		}\bibfield  {title} {\bibinfo {title} {Applications of {DFT} + {DMFT} in
				materials science},\ }\href
		{https://doi.org/10.1146/annurev-matsci-070218-121825} {\bibfield  {journal}
			{\bibinfo  {journal} {Annu. Rev. Mater. Res.}\ }\textbf {\bibinfo {volume}
				{49}},\ \bibinfo {pages} {31} (\bibinfo {year} {2019})}\BibitemShut {NoStop}%
		\bibitem [{\citenamefont {Kotliar}\ \emph {et~al.}(2006)\citenamefont
			{Kotliar}, \citenamefont {Savrasov}, \citenamefont {Haule}, \citenamefont
			{Oudovenko}, \citenamefont {Parcollet},\ and\ \citenamefont
			{Marianetti}}]{kotliar_electronic_2006}%
		\BibitemOpen
		\bibfield  {author} {\bibinfo {author} {\bibfnamefont {G.}~\bibnamefont
				{Kotliar}}, \bibinfo {author} {\bibfnamefont {S.~Y.}\ \bibnamefont
				{Savrasov}}, \bibinfo {author} {\bibfnamefont {K.}~\bibnamefont {Haule}},
			\bibinfo {author} {\bibfnamefont {V.~S.}\ \bibnamefont {Oudovenko}}, \bibinfo
			{author} {\bibfnamefont {O.}~\bibnamefont {Parcollet}},\ and\ \bibinfo
			{author} {\bibfnamefont {C.~A.}\ \bibnamefont {Marianetti}},\ }\bibfield
		{title} {\bibinfo {title} {Electronic structure calculations with dynamical
				mean-field theory},\ }\href
		{https://link.aps.org/doi/10.1103/RevModPhys.78.865} {\bibfield  {journal}
			{\bibinfo  {journal} {Rev. Mod. Phys.}\ }\textbf {\bibinfo {volume} {78}},\
			\bibinfo {pages} {865} (\bibinfo {year} {2006})}\BibitemShut {NoStop}%
		\bibitem [{\citenamefont {Zhou}\ \emph {et~al.}(2021)\citenamefont {Zhou},
			\citenamefont {Pandey},\ and\ \citenamefont {Feng}}]{zhou_dynamical_2021}%
		\BibitemOpen
		\bibfield  {author} {\bibinfo {author} {\bibfnamefont {Z.}~\bibnamefont
				{Zhou}}, \bibinfo {author} {\bibfnamefont {S.~K.}\ \bibnamefont {Pandey}},\
			and\ \bibinfo {author} {\bibfnamefont {J.}~\bibnamefont {Feng}},\ }\bibfield
		{title} {\bibinfo {title} {Dynamical correlation enhanced orbital magnetization in VI$_3$},\ }\href
		{https://link.aps.org/doi/10.1103/PhysRevB.103.035137} {\bibfield  {journal}
			{\bibinfo  {journal} {Phys. Rev. B}\ }\textbf {\bibinfo {volume} {103}},\
			\bibinfo {pages} {035137} (\bibinfo {year} {2021})}\BibitemShut {NoStop}%
		\bibitem [{\citenamefont {Kvashnin}\ \emph {et~al.}(2022)\citenamefont
			{Kvashnin}, \citenamefont {Rudenko}, \citenamefont {Thunström},
			\citenamefont {Rösner},\ and\ \citenamefont
			{Katsnelson}}]{kvashnin_dynamical_2022}%
		\BibitemOpen
		\bibfield  {author} {\bibinfo {author} {\bibfnamefont {Y.~O.}\ \bibnamefont
				{Kvashnin}}, \bibinfo {author} {\bibfnamefont {A.~N.}\ \bibnamefont
				{Rudenko}}, \bibinfo {author} {\bibfnamefont {P.}~\bibnamefont {Thunström}},
			\bibinfo {author} {\bibfnamefont {M.}~\bibnamefont {Rösner}},\ and\ \bibinfo
			{author} {\bibfnamefont {M.~I.}\ \bibnamefont {Katsnelson}},\ }\bibfield
		{title} {\bibinfo {title} {Dynamical correlations in single-layer
				{CrI$_3$}},\ }\href {https://link.aps.org/doi/10.1103/PhysRevB.105.205124}
		{\bibfield  {journal} {\bibinfo  {journal} {Phys. Rev. B}\ }\textbf {\bibinfo
				{volume} {105}},\ \bibinfo {pages} {205124} (\bibinfo {year}
			{2022})}\BibitemShut {NoStop}%
		\bibitem [{\citenamefont {Maiti}\ \emph {et~al.}(2004)\citenamefont {Maiti},
			\citenamefont {Kumar}, \citenamefont {Sarma}, \citenamefont {Weschke},\ and\
			\citenamefont {Kaindl}}]{LaCaVO}%
		\BibitemOpen
		\bibfield  {author} {\bibinfo {author} {\bibfnamefont {K.}~\bibnamefont
				{Maiti}}, \bibinfo {author} {\bibfnamefont {A.}~\bibnamefont {Kumar}},
			\bibinfo {author} {\bibfnamefont {D.~D.}\ \bibnamefont {Sarma}}, \bibinfo
			{author} {\bibfnamefont {E.}~\bibnamefont {Weschke}},\ and\ \bibinfo {author}
			{\bibfnamefont {G.}~\bibnamefont {Kaindl}},\ }\bibfield  {title} {\bibinfo
			{title} {Surface and bulk electronic structure of {La$_{1-x}$Ca$_x$VO$_3$}},\
		}\href {https://link.aps.org/doi/10.1103/PhysRevB.70.195112} {\bibfield
			{journal} {\bibinfo  {journal} {Phys. Rev. B}\ }\textbf {\bibinfo {volume}
				{70}},\ \bibinfo {pages} {195112} (\bibinfo {year} {2004})}\BibitemShut
		{NoStop}%
		\bibitem [{\citenamefont {Maiti}\ \emph {et~al.}(2005)\citenamefont {Maiti},
			\citenamefont {Singh}, \citenamefont {Medicherla}, \citenamefont {Rayaprol},\
			and\ \citenamefont {Sampathkumaran}}]{BaIrO3}%
		\BibitemOpen
		\bibfield  {author} {\bibinfo {author} {\bibfnamefont {K.}~\bibnamefont
				{Maiti}}, \bibinfo {author} {\bibfnamefont {R.~S.}\ \bibnamefont {Singh}},
			\bibinfo {author} {\bibfnamefont {V.~R.~R.}\ \bibnamefont {Medicherla}},
			\bibinfo {author} {\bibfnamefont {S.}~\bibnamefont {Rayaprol}},\ and\
			\bibinfo {author} {\bibfnamefont {E.~V.}\ \bibnamefont {Sampathkumaran}},\
		}\bibfield  {title} {\bibinfo {title} {Origin of charge density wave
				formation in insulators from a high resolution photoemission study of
				{BaIrO$_{3}$}},\ }\href@noop {} {\bibfield  {journal} {\bibinfo  {journal}
				{Phys. Rev. Lett.}\ }\textbf {\bibinfo {volume} {95}},\ \bibinfo {pages}
			{016404} (\bibinfo {year} {2005})}\BibitemShut {NoStop}%
		\bibitem [{\citenamefont {Bansal}\ \emph {et~al.}(2023)\citenamefont {Bansal},
			\citenamefont {Maurya}, \citenamefont {Ali}, \citenamefont {Reddy},\ and\
			\citenamefont {Singh}}]{LSNO}%
		\BibitemOpen
		\bibfield  {author} {\bibinfo {author} {\bibfnamefont {S.}~\bibnamefont
				{Bansal}}, \bibinfo {author} {\bibfnamefont {R.~K.}\ \bibnamefont {Maurya}},
			\bibinfo {author} {\bibfnamefont {A.}~\bibnamefont {Ali}}, \bibinfo {author}
			{\bibfnamefont {B.~H.}\ \bibnamefont {Reddy}},\ and\ \bibinfo {author}
			{\bibfnamefont {R.~S.}\ \bibnamefont {Singh}},\ }\bibfield  {title} {\bibinfo
			{title} {Role of electron correlation and disorder on the electronic
				structure of layered nickelate
				${({\mathrm{La}}_{0.5}{\mathrm{Sr}}_{0.5})}_{2} ${N}i{O}$_4$},\ }\href
		{https://link.aps.org/doi/10.1103/PhysRevMaterials.7.064007} {\bibfield
			{journal} {\bibinfo  {journal} {Phys. Rev. Materials}\ }\textbf {\bibinfo
				{volume} {7}},\ \bibinfo {pages} {064007} (\bibinfo {year}
			{2023})}\BibitemShut {NoStop}%
		\bibitem [{\citenamefont {Norman}\ \emph {et~al.}(1998)\citenamefont {Norman},
			\citenamefont {Ding}, \citenamefont {Randeria}, \citenamefont {Campuzano},
			\citenamefont {Yokoya}, \citenamefont {Takeuchi}, \citenamefont {Takahashi},
			\citenamefont {Mochiku}, \citenamefont {Kadowaki}, \citenamefont
			{Guptasarma},\ and\ \citenamefont {Hinks}}]{norman_destruction_1998}%
		\BibitemOpen
		\bibfield  {author} {\bibinfo {author} {\bibfnamefont {M.~R.}\ \bibnamefont
				{Norman}}, \bibinfo {author} {\bibfnamefont {H.}~\bibnamefont {Ding}},
			\bibinfo {author} {\bibfnamefont {M.}~\bibnamefont {Randeria}}, \bibinfo
			{author} {\bibfnamefont {J.~C.}\ \bibnamefont {Campuzano}}, \bibinfo {author}
			{\bibfnamefont {T.}~\bibnamefont {Yokoya}}, \bibinfo {author} {\bibfnamefont
				{T.}~\bibnamefont {Takeuchi}}, \bibinfo {author} {\bibfnamefont
				{T.}~\bibnamefont {Takahashi}}, \bibinfo {author} {\bibfnamefont
				{T.}~\bibnamefont {Mochiku}}, \bibinfo {author} {\bibfnamefont
				{K.}~\bibnamefont {Kadowaki}}, \bibinfo {author} {\bibfnamefont
				{P.}~\bibnamefont {Guptasarma}},\ and\ \bibinfo {author} {\bibfnamefont
				{D.~G.}\ \bibnamefont {Hinks}},\ }\bibfield  {title} {\bibinfo {title} {Destruction of the {Fermi} surface in underdoped
				high-Tc superconductors},\ }\href {https://www.nature.com/articles/32366}
		{\bibfield  {journal} {\bibinfo  {journal} {Nature}\ }\textbf {\bibinfo
				{volume} {392}},\ \bibinfo {pages} {157} (\bibinfo {year}
			{1998})}\BibitemShut {NoStop}%
		\bibitem [{\citenamefont {Singh}\ \emph {et~al.}(2008)\citenamefont {Singh},
			\citenamefont {Medicherla}, \citenamefont {Maiti},\ and\ \citenamefont
			{Sampathkumaran}}]{Y2Ir2O7}%
		\BibitemOpen
		\bibfield  {author} {\bibinfo {author} {\bibfnamefont {R.~S.}\ \bibnamefont
				{Singh}}, \bibinfo {author} {\bibfnamefont {V.~R.~R.}\ \bibnamefont
				{Medicherla}}, \bibinfo {author} {\bibfnamefont {K.}~\bibnamefont {Maiti}},\
			and\ \bibinfo {author} {\bibfnamefont {E.~V.}\ \bibnamefont
				{Sampathkumaran}},\ }\bibfield  {title} {\bibinfo {title} {Evidence for
				strong $5d$ electron correlations in the pyrochlore {Y$_2$I}r$_2${O}$_7$
				studied using high-resolution photoemission spectroscopy},\ }\href
		{https://link.aps.org/doi/10.1103/PhysRevB.77.201102} {\bibfield  {journal}
			{\bibinfo  {journal} {Phys. Rev. B}\ }\textbf {\bibinfo {volume} {77}},\
			\bibinfo {pages} {201102} (\bibinfo {year} {2008})}\BibitemShut {NoStop}%
		\bibitem [{\citenamefont {Reddy}\ \emph {et~al.}(2019)\citenamefont {Reddy},
			\citenamefont {Ali},\ and\ \citenamefont {Singh}}]{reddy_role_2019}%
		\BibitemOpen
		\bibfield  {author} {\bibinfo {author} {\bibfnamefont {B.~H.}\ \bibnamefont
				{Reddy}}, \bibinfo {author} {\bibfnamefont {A.}~\bibnamefont {Ali}},\ and\
			\bibinfo {author} {\bibfnamefont {R.~S.}\ \bibnamefont {Singh}},\ }\bibfield
		{title} {\bibinfo {title} {Role of disorder and strong 5$d$ electron
				correlation in the electronic structure of {Sr$_2$TiIrO$_6$}},\ }\href
		{https://doi.org/10.1209/0295-5075/127/47003} {\bibfield  {journal} {\bibinfo
				{journal} {Europhysics Letters}\ }\textbf {\bibinfo {volume} {127}},\
			\bibinfo {pages} {47003} (\bibinfo {year} {2019})}\BibitemShut {NoStop}%
		\bibitem [{\citenamefont {Hegger}\ \emph {et~al.}(2000)\citenamefont {Hegger},
			\citenamefont {Petrovic}, \citenamefont {Moshopoulou}, \citenamefont
			{Hundley}, \citenamefont {Sarrao}, \citenamefont {Fisk},\ and\ \citenamefont
			{Thompson}}]{hegger_pressure-induced_2000}%
		\BibitemOpen
		\bibfield  {author} {\bibinfo {author} {\bibfnamefont {H.}~\bibnamefont
				{Hegger}}, \bibinfo {author} {\bibfnamefont {C.}~\bibnamefont {Petrovic}},
			\bibinfo {author} {\bibfnamefont {E.~G.}\ \bibnamefont {Moshopoulou}},
			\bibinfo {author} {\bibfnamefont {M.~F.}\ \bibnamefont {Hundley}}, \bibinfo
			{author} {\bibfnamefont {J.~L.}\ \bibnamefont {Sarrao}}, \bibinfo {author}
			{\bibfnamefont {Z.}~\bibnamefont {Fisk}},\ and\ \bibinfo {author}
			{\bibfnamefont {J.~D.}\ \bibnamefont {Thompson}},\ }\bibfield  {title}
		{\bibinfo {title} {Pressure-{Induced} {Superconductivity} in {Quasi}-{2D}
				{CeRhIn}$_5$},\ }\href {https://link.aps.org/doi/10.1103/PhysRevLett.84.4986}
		{\bibfield  {journal} {\bibinfo  {journal} {Phys. Rev. Lett.}\ }\textbf
			{\bibinfo {volume} {84}},\ \bibinfo {pages} {4986} (\bibinfo {year}
			{2000})}\BibitemShut {NoStop}%
		\bibitem [{\citenamefont {Krellner}\ \emph {et~al.}(2008)\citenamefont
			{Krellner}, \citenamefont {Förster}, \citenamefont {Jeevan}, \citenamefont
			{Geibel},\ and\ \citenamefont {Sichelschmidt}}]{krellner_relevance_2008}%
		\BibitemOpen
		\bibfield  {author} {\bibinfo {author} {\bibfnamefont {C.}~\bibnamefont
				{Krellner}}, \bibinfo {author} {\bibfnamefont {T.}~\bibnamefont {Förster}},
			\bibinfo {author} {\bibfnamefont {H.}~\bibnamefont {Jeevan}}, \bibinfo
			{author} {\bibfnamefont {C.}~\bibnamefont {Geibel}},\ and\ \bibinfo {author}
			{\bibfnamefont {J.}~\bibnamefont {Sichelschmidt}},\ }\bibfield  {title}
		{\bibinfo {title} {Relevance of {Ferromagnetic} {Correlations} for the
				{Electron} {Spin} {Resonance} in {Kondo} {Lattice} {Systems}},\ }\href
		{https://link.aps.org/doi/10.1103/PhysRevLett.100.066401} {\bibfield
			{journal} {\bibinfo  {journal} {Phys. Rev. Lett.}\ }\textbf {\bibinfo
				{volume} {100}},\ \bibinfo {pages} {066401} (\bibinfo {year}
			{2008})}\BibitemShut {NoStop}%
		\bibitem [{\citenamefont {Fisk}\ \emph {et~al.}(1988)\citenamefont {Fisk},
			\citenamefont {Hess}, \citenamefont {Pethick}, \citenamefont {Pines},
			\citenamefont {Smith}, \citenamefont {Thompson},\ and\ \citenamefont
			{Willis}}]{fisk_heavy-electron_1988}%
		\BibitemOpen
		\bibfield  {author} {\bibinfo {author} {\bibfnamefont {Z.}~\bibnamefont
				{Fisk}}, \bibinfo {author} {\bibfnamefont {D.~W.}\ \bibnamefont {Hess}},
			\bibinfo {author} {\bibfnamefont {C.~J.}\ \bibnamefont {Pethick}}, \bibinfo
			{author} {\bibfnamefont {D.}~\bibnamefont {Pines}}, \bibinfo {author}
			{\bibfnamefont {J.~L.}\ \bibnamefont {Smith}}, \bibinfo {author}
			{\bibfnamefont {J.~D.}\ \bibnamefont {Thompson}},\ and\ \bibinfo {author}
			{\bibfnamefont {J.~O.}\ \bibnamefont {Willis}},\ }\bibfield  {title}
		{\bibinfo {title} {Heavy-{Electron} {Metals}: {New} {Highly} {Correlated}
				{States} of {Matter}},\ }\href
		{https://www.science.org/doi/10.1126/science.239.4835.33} {\bibfield
			{journal} {\bibinfo  {journal} {Science}\ }\textbf {\bibinfo {volume}
				{239}},\ \bibinfo {pages} {33} (\bibinfo {year} {1988})}\BibitemShut
		{NoStop}%
		\bibitem [{\citenamefont {Mravlje}\ \emph {et~al.}(2011)\citenamefont
			{Mravlje}, \citenamefont {Aichhorn}, \citenamefont {Miyake}, \citenamefont
			{Haule}, \citenamefont {Kotliar},\ and\ \citenamefont
			{Georges}}]{mravlje_coherence-incoherence_2011}%
		\BibitemOpen
		\bibfield  {author} {\bibinfo {author} {\bibfnamefont {J.}~\bibnamefont
				{Mravlje}}, \bibinfo {author} {\bibfnamefont {M.}~\bibnamefont {Aichhorn}},
			\bibinfo {author} {\bibfnamefont {T.}~\bibnamefont {Miyake}}, \bibinfo
			{author} {\bibfnamefont {K.}~\bibnamefont {Haule}}, \bibinfo {author}
			{\bibfnamefont {G.}~\bibnamefont {Kotliar}},\ and\ \bibinfo {author}
			{\bibfnamefont {A.}~\bibnamefont {Georges}},\ }\bibfield  {title} {\bibinfo
			{title} {Coherence-{Incoherence} {Crossover} and the {Mass}-{Renormalization}
				{Puzzles} in {Sr$_2$RuO$_4$}},\ }\href
		{https://link.aps.org/doi/10.1103/PhysRevLett.106.096401} {\bibfield
			{journal} {\bibinfo  {journal} {Phys. Rev. Lett.}\ }\textbf {\bibinfo
				{volume} {106}},\ \bibinfo {pages} {096401} (\bibinfo {year}
			{2011})}\BibitemShut {NoStop}%
		\bibitem [{\citenamefont {Kvashnin}\ \emph {et~al.}(2015)\citenamefont
			{Kvashnin}, \citenamefont {Grånäs}, \citenamefont {Di~Marco}, \citenamefont
			{Katsnelson}, \citenamefont {Lichtenstein},\ and\ \citenamefont
			{Eriksson}}]{kvashnin_exchange_2015}%
		\BibitemOpen
		\bibfield  {author} {\bibinfo {author} {\bibfnamefont {Y.~O.}\ \bibnamefont
				{Kvashnin}}, \bibinfo {author} {\bibfnamefont {O.}~\bibnamefont {Grånäs}},
			\bibinfo {author} {\bibfnamefont {I.}~\bibnamefont {Di~Marco}}, \bibinfo
			{author} {\bibfnamefont {M.~I.}\ \bibnamefont {Katsnelson}}, \bibinfo
			{author} {\bibfnamefont {A.~I.}\ \bibnamefont {Lichtenstein}},\ and\ \bibinfo
			{author} {\bibfnamefont {O.}~\bibnamefont {Eriksson}},\ }\bibfield  {title}
		{\bibinfo {title} {Exchange parameters of strongly correlated materials:
				{Extraction} from spin-polarized density functional theory plus dynamical
				mean-field theory},\ }\href
		{https://link.aps.org/doi/10.1103/PhysRevB.91.125133} {\bibfield  {journal}
			{\bibinfo  {journal} {Phys. Rev. B}\ }\textbf {\bibinfo {volume} {91}},\
			\bibinfo {pages} {125133} (\bibinfo {year} {2015})}\BibitemShut {NoStop}%
		\bibitem [{\citenamefont {Mizutani}(2001)}]{Mizutani}%
		\BibitemOpen
		\bibfield  {author} {\bibinfo {author} {\bibfnamefont {U.}~\bibnamefont
				{Mizutani}},\ }\href {https://books.google.co.in/books?id=zY5z_UGqAcwC}
		{\emph {\bibinfo {title} {Introduction to the Electron Theory of Metals}}}\
		(\bibinfo  {publisher} {Cambridge University Press},\ \bibinfo {year}
		{2001})\BibitemShut {NoStop}%
	\end{thebibliography}
	%\iffalse
	
\end{document}

\title{Supplemental Material for ``Manifestation of incoherent-coherent crossover and non-Stoner magnetism in the electronic structure of Fe$_3$GeTe$_2$"}% Force line breaks with \\
	
\author{Deepali Sharma}
%\email{deepali19@iiserb.ac.in}
\affiliation{Department of Physics, Indian Institute of Science Education and Research Bhopal, Bhopal Bypass Road, Bhauri, Bhopal 462066, India}%
\author{Asif Ali}
\affiliation{Department of Physics, Indian Institute of Science Education and Research Bhopal, Bhopal Bypass Road, Bhauri, Bhopal 462066, India}%
\author{Neeraj Bhatt}
\affiliation{Department of Physics, Indian Institute of Science Education and Research Bhopal, Bhopal Bypass Road, Bhauri, Bhopal 462066, India}%
\author{Rajeswari Roy Chowdhury}
\affiliation{Department of Physics, Indian Institute of Science Education and Research Bhopal, Bhopal Bypass Road, Bhauri, Bhopal 462066, India}%
\author{Chandan Patra}
\affiliation{Department of Physics, Indian Institute of Science Education and Research Bhopal, Bhopal Bypass Road, Bhauri, Bhopal 462066, India}%
\author{Ravi Prakash Singh}
\affiliation{Department of Physics, Indian Institute of Science Education and Research Bhopal, Bhopal Bypass Road, Bhauri, Bhopal 462066, India}%
\author{Ravi Shankar Singh}
\email{rssingh@iiserb.ac.in}
\affiliation{Department of Physics, Indian Institute of Science Education and Research Bhopal, Bhopal Bypass Road, Bhauri, Bhopal 462066, India}%
		
\date{\today}% It is always \today, today,

\maketitle

\section*{\label{sec:level1} Note 1: Experimental Details and Analysis}

\begin{figure}[H]
	\centering
	\includegraphics[width=.79\textwidth]{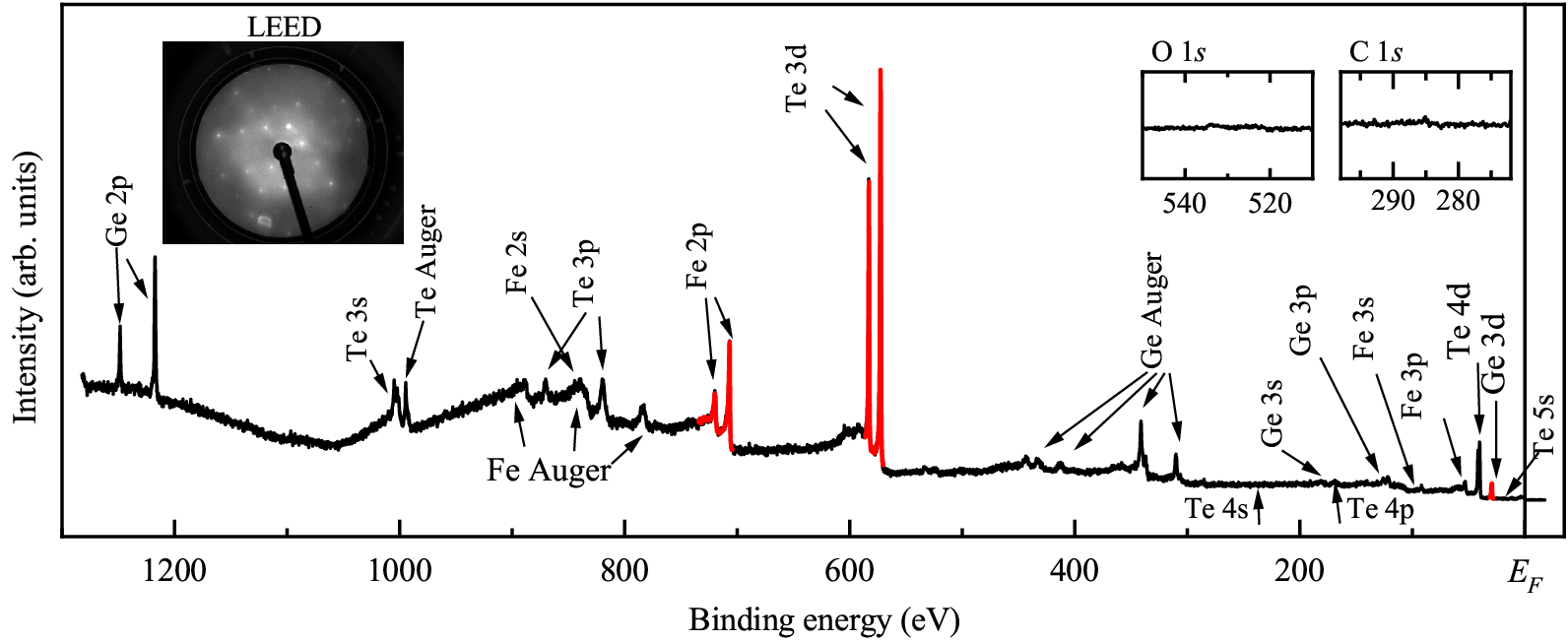}
	\caption{Survey scan of Fe$_{3}$GeTe$_{2}$ collected using Al $K_{\alpha}$ radiation at 30 K. All the core level and Auger features are marked by arrows. The LEED pattern exhibiting hexagonal symmetry of the spots is shown in the inset on the left. Two insets on the right show the absence of O 1\textit{s} and C 1\textit{s} features in the corresponding binding energy regions.} \label{fig:SFig1}
\end{figure}

\textbf{X-ray photoemission survey scan} : \\
The survey scan collected using Al $K_{\alpha}$ radiation at 30 K is shown in Fig. \ref{fig:SFig1}. The Fe 2$p$, Te 3$d$ and Ge 3$d$ core levels are marked with red colour which are also shown in the Fig. 1(b) of the main text. Absence of O 1\textit{s} and C 1\textit{s} features in the spectra as shown in the insets, along with bright spots in the LEED pattern exhibiting hexagonal symmetry, confirms high quality and clean sample surface.

\begin{figure}[H]
	\centering
	\includegraphics[width=.4\textwidth]{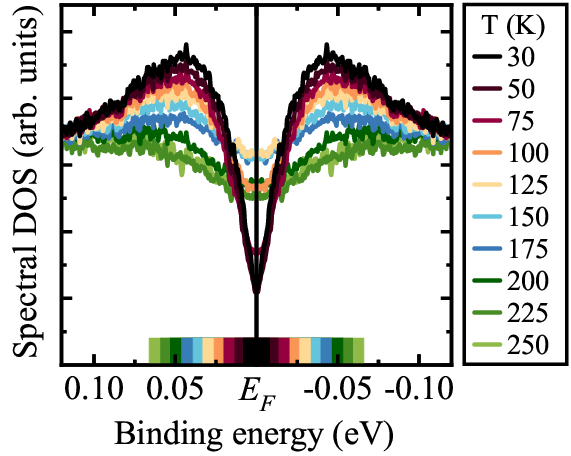}
	\caption{ Symmetrization of the high-resolution He {\scriptsize I} spectra in the vicinity of $E_{F}$ for different temperature. Bottom color bar indicates the thermal energy ($\pm~3k_BT$) range.} \label{fig:SFig2}
\end{figure}

\pagebreak
\textbf{Temperature dependent high-resolution spectral density of states (SDOS)}: \\
     The photoemission intensity can be expressed as $I(E)= DOS(E,T) * \Gamma ^e * \Gamma ^h * G(E) * F(E,T)$, where $\Gamma ^e$ ($\Gamma ^h$) represents lifetime broadening of photoelectron (photohole), \textit{F} and \textit{G} represent Fermi-Dirac (FD) distribution function and instrumental resolution broadening, respectively \cite{LaCaVO,*BaIrO3,*LSNO}. Since the photoelectrons and photoholes are long-lived close to the Fermi energy, $E_{F}$, the photoemission intensity divided by the resolution broadened FD function depicts the spectral DOS (SDOS) [$I(E)/(G(E) * F(E,T))$]. SDOS obtained using this method is shown in the main text (Fig. 3(b)). Since the FD distribution function (also resolution broadened) follows the relation \textit{F(E)}+\textit{F(-E)} = 1, thus symmetrization of the photoemmision spectra around $E_{F}$ [\textit{I(E)}+\textit{I(-E)}] also provides a good description of SDOS, assuming DOS to be a smooth function around $E_{F}$. The SDOS obtained by this method (symmetrization) has been shown in Fig. \ref{fig:SFig2}, clearly indicating emergence of the quasiparticle peak ($\sim$40 meV) and the peculiar behavior of SDOS at $E_{F}$, same as the results shown in Fig. 3(b) and (c) of the main text.
     
\section*{\label{sec:level2} Note 2: Computational Details and Analysis}     

\begin{figure*}[!b]
	\centering
	\includegraphics[width=0.55\textwidth]{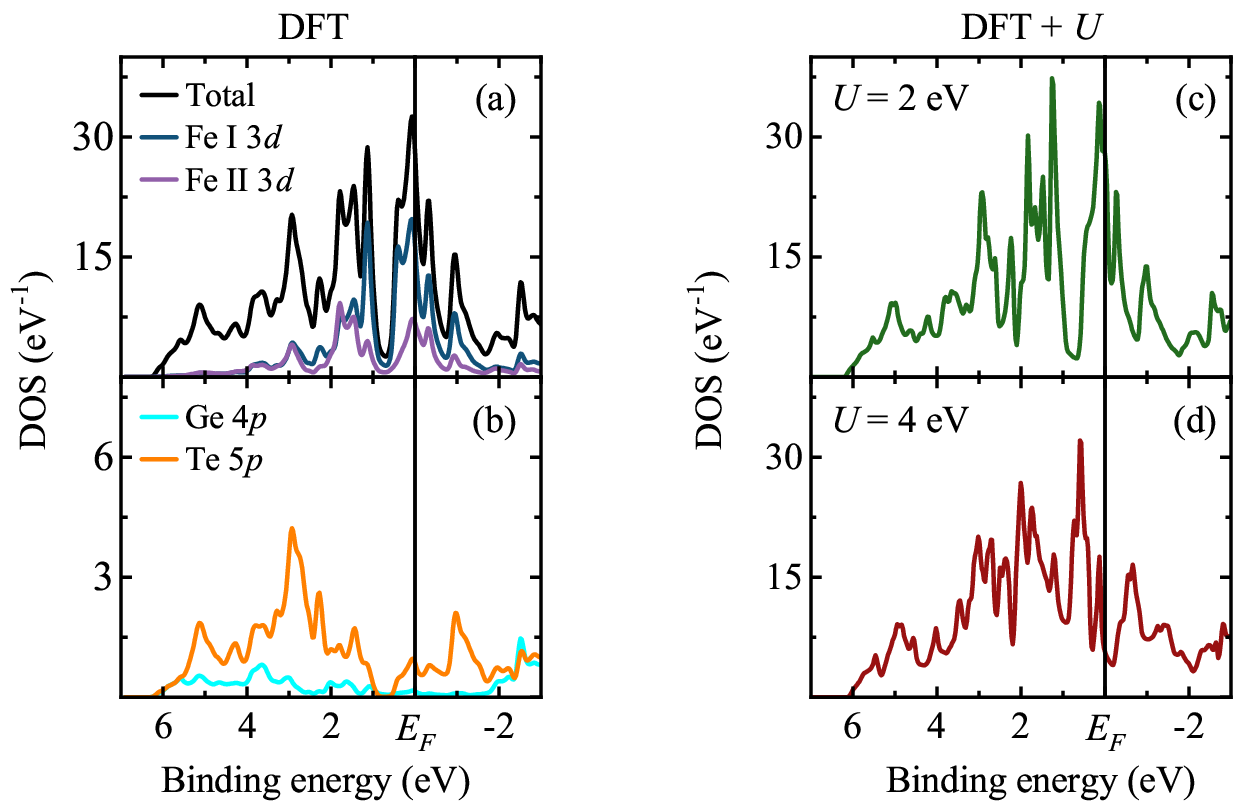}
	\caption{(a) Total DOS and partial DOSs of Fe I 3\textit{d} and Fe II 3\textit{d}, and  (b) partial DOSs of Ge 4\textit{p} and Te 5\textit{p} obtained from NM DFT calculation. Total DOS obtained from NM DFT+\textit{U} calculations (c) for $U$ = 2 eV, and (d) for $U$ = 4 eV.} \label{fig:SFig3}
\end{figure*}

\textbf{Non-magnetic DFT and DFT +$U$ results} :\\
Figure \ref{fig:SFig3} shows the total density of states (DOS) and partial DOSs of Fe I 3$d$, Fe II 3$d$, Ge 4$p$ and Te 5$p$ obtained from the non-magnetic (NM) DFT calculation. The partial DOSs show contributions from inequivalent Fe within $\pm$ 2 eV binding energy (BE) while the dominant contributions from Te 5\textit{p} and Ge 4\textit{p} states appear above 2 eV BE, in agreement with experimental valence band spectra. Total DOS obtained from DFT+$U$ calculations ($U$ = 2 eV and 4 eV) are shown in the right panels of Fig. \ref{fig:SFig3}. The inclusion of \textit{U} in DFT leads to shift of 1.5 eV feature towards higher BE and decrease in the DOS at $E_{F}$, in contrast to the experimental spectra where the Fe 3\textit{d} band appears within the 1.5 eV BE.

\begin{figure}
	\centering
	\includegraphics[width=0.7\textwidth]{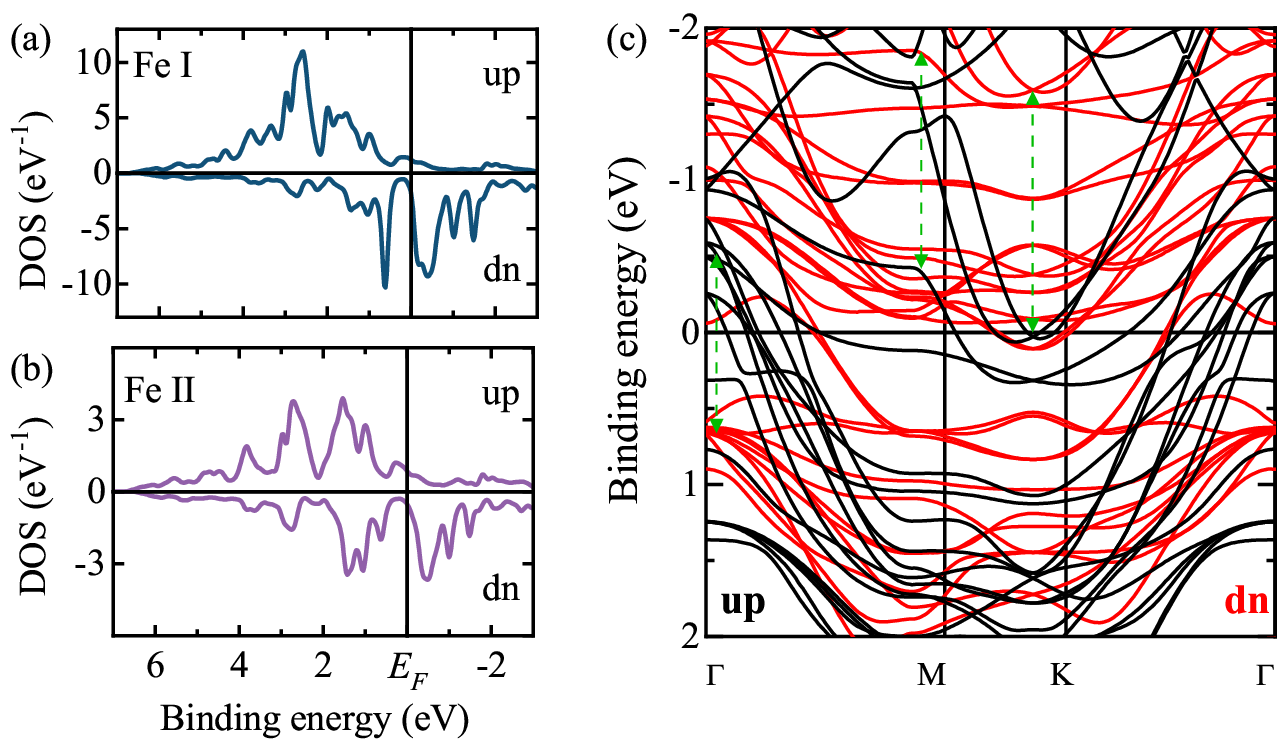}
	\caption{ Spin-polarized partial DOSs for (a) Fe I and (b) Fe II using FM DFT calculation. (c) Spin-polarized band dispersions along high symmetry directions using FM DFT calculation. The vertical green dashed lines are guide to the eyes for approximate exchange splitting of bands.}\label{fig:SFig4}
\end{figure}

\begin{figure*}[!b]
	\centering
	\includegraphics[width=0.85\textwidth]{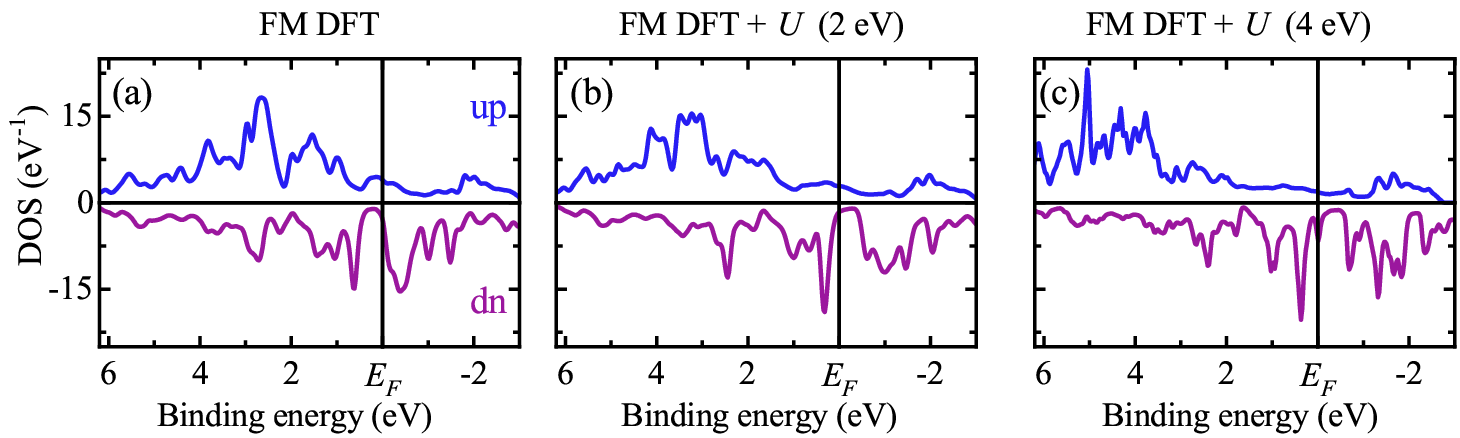}
	\caption{ Spin-polarized total DOS using (a) FM DFT, (b) FM DFT+$U$ (2 eV) and (c) FM DFT+$U$ (4 eV), calculations.} \label{fig:SFig5}
\end{figure*}

\textbf{Ferromagnetic DFT and DFT+$U$ results} : \\
Results of the ferromagnetic (FM) DFT calculations are shown in the Fig. \ref{fig:SFig4} where, spin-polarized 3\textit{d} states corresponding to Fe I and Fe II are shown in panels (a) and (b), respectively. Large exchange splitting of spin-polarized DOS is clearly evident for both the inequivalent Fe alongwith dominant contributions of 3\textit{d} states appearing beyond 2 eV BE. FM DFT calculation also overestimates spin magnetic moments (2.46 $\mu_{B}$ and 1.59 $\mu_{B}$ for Fe I and Fe II, respectively), in comparison to the experimental result \cite{chowdhury_unconventional_2021}. DOS($E_{F}$) = 3.31 states eV$^{-1}$ f.u.$^{-1}$ leads to Sommerfeld coefficient ($\gamma_{_{\text{DFT}}}$ = ${\pi}$$^{2}$$k_B$$^{2}$DOS($E_{F}$)/3) of 7.79 mJ mol$^{-1}$ K$^{-2}$. With the experimentally obtained large ${\gamma}$ value for Fe$_{3}$GeTe$_{2}$ \cite{zhu_electronic_2016}, the FM DFT calculation suggests a significant quasiparticle mass enhancement $m^{*}$/$m_{_{\text{DFT}}}$ = ${\gamma}$/$\gamma_{_{\text{DFT}}}$ = 14.12. Spin-polarised band dispersion along high symmetry directions are shown in Fig. \ref{fig:SFig4} (c) where, the dashed vertical lines show approximate exchange splittings of bands. The spin-polarised total DOS obtained within FM DFT and FM DFT+$U$ for $U$ = 2 eV and $U$ = 4 eV are shown in Fig. \ref{fig:SFig5} (a)-(c). The inclusion of $U$ in FM DFT calculations lead to further enhancement of exchange splitting and magnetic moment. For $U$ = 2 eV and 4 eV, the obtained spin magnetic moments are 2.79 $\mu_{B}$ (1.76 $\mu_{B}$) and 3.08 $\mu_{B}$ (2.38 $\mu_{B}$) for Fe I (Fe II), respectively.

